\documentclass[11pt]{article}

\usepackage[margin=1in]{geometry}
\usepackage{lmodern}
\usepackage[T1]{fontenc}
\usepackage[utf8]{inputenc}
\usepackage{amsmath,amssymb,amsthm,mathtools}
\usepackage{mathrsfs}
\usepackage{booktabs}
\usepackage{enumitem}
\usepackage[colorlinks=true,linkcolor=blue,citecolor=blue,urlcolor=blue]{hyperref}
\usepackage{microtype}
\usepackage{xcolor}
\usepackage{tikz}
\usetikzlibrary{decorations.pathmorphing,positioning}
\usepackage{array}

\usepackage[backend=bibtex,style=alphabetic,maxnames=99]{biblatex}
\numberwithin{equation}{section}

\theoremstyle{plain}
\newtheorem{theorem}{Theorem}[section]
\newtheorem{lemma}[theorem]{Lemma}
\newtheorem{proposition}[theorem]{Proposition}
\newtheorem{corollary}[theorem]{Corollary}
\newtheorem{conjecture}[theorem]{Conjecture}

\theoremstyle{definition}
\newtheorem{definition}[theorem]{Definition}

\theoremstyle{remark}
\newtheorem{remark}[theorem]{Remark}

\newcommand{\Adv}{\operatorname{Adv}}
\newcommand{\F}{\mathbb{F}_2}
\newcommand{\N}{\mathbb{N}}
\newcommand{\C}{\mathbb{C}}
\newcommand{\cH}{\mathcal{H}}
\newcommand{\id}{\mathrm{id}}
\newcommand{\bx}{\mathbf{x}}
\newcommand{\ba}{\mathbf{a}}
\newcommand{\alg}{\mathfrak{A}}

\DeclareMathOperator{\sat}{sat}
\DeclareMathOperator{\qsat}{qsat}
\DeclareMathOperator{\dist}{dist}
\DeclareMathOperator{\tr}{tr}
\DeclareMathOperator{\ar}{ar}
\DeclareMathOperator{\PVM}{PVM}
\DeclareMathOperator{\spn}{span}

\DeclareMathOperator{\supp}{supp}
\DeclareMathOperator{\enc}{enc}
\DeclareMathOperator{\Comp}{Comp}
\DeclareMathOperator{\val}{val}
\DeclareMathOperator{\Triples}{Triples}

\newcommand{\cX}{\mathcal{X}}
\newcommand{\cL}{\mathcal{L}}
\newcommand{\cW}{\mathcal{W}}
\newcommand{\cR}{\mathcal{R}}
\newcommand{\cE}{\mathcal{E}}
\newcommand{\cU}{\mathcal{U}}
\newcommand{\cA}{\mathcal{A}}
\newcommand{\cV}{\mathcal{V}}
\newcommand{\cN}{\mathcal{N}}
\newcommand{\sE}{\mathscr{E}}
\newcommand{\sA}{\mathscr{A}}
\newcommand{\Jin}{\mathcal{J}_{\mathrm{in}}}
\newcommand{\Jmid}{\mathcal{J}_{\mathrm{mid}}}
\newcommand{\Thin}{\mathrm{Thin}}
\newcommand{\GameThree}{\mathsf{Game3}}
\newcommand{\PhiF}{\Phi_{\mathrm{FMW}}}
\newcommand{\oplusU}[1]{{\textstyle\bigoplus}#1}

\newcommand{\empt}{\mathtt{empty}}

\title{On the quantum chromatic gap II}

\author{Lorenzo Ciardo\\
Graz University of Technology\\
\texttt{lorenzo.ciardo@tugraz.at}
\and
Antoine Mottet\\
Hamburg University of Technology\\
\texttt{antoine.mottet@tuhh.de}
}

\date{}

\begin{document}

\maketitle

\begin{abstract}
\noindent We confirm a conjecture of the first author [SODA'26] on a strong form of
pseudo-telepathy for Khot's $d$-to-$1$ games: for every $d\geq 4$, every $\epsilon>0$, and every $k\in\N$, there is a $d$-to-$1$ label-cover game of classical value
less than $\epsilon$ that admits a perfect quantum strategy
using finite-dimensional shared entanglement with measurement contexts of radius at least $k$. As a consequence, there exist
graphs whose quantum chromatic number $\chi_q$ is $3$ and whose classical chromatic
number $\chi$ is arbitrarily large.
This is the first super-exponential 
$\chi_q$ vs. $\chi$
gap, and it improves on the exponential lower bounds of Brassard, Cleve, and Tapp [Phys.\ Rev.\ Lett.\ '99], who produced graphs for which $\chi\geq 1.1397^{\chi_q}$, and Lalonde, who recently achieved $\chi\geq 1.1546^{\chi_q}$.

To prove the conjecture, we proceed as follows. 
We start with a ``stretched'' version of the Mermin--Peres magic square, obtained by spacing out its equations and linking them with additional long equation wires. This operation increases the radius of the initial measurement contexts of a perfect quantum strategy for the game, while preserving classical unsatisfiability. We then apply parallel repetition to make the classical value small, and we turn the system into a pseudo-telepathic
multi-layered Kneser hypergraph via a PCP construction of Dinur, Guruswami, Khot, and Regev [STOC'03]. Finally, we feed the resulting hypergraph into the reduction of Fei, Minzer, and Wang from hypergraph colouring to
$4$-to-$1$ games, we establish quantum completeness of this reduction, and we show that it preserves context radius.  
\end{abstract}

\setcounter{page}{0}\thispagestyle{empty}\clearpage

\newpage
\tableofcontents

\setcounter{page}{0}\thispagestyle{empty}\clearpage

\section{Introduction}\label{sec:intro}

Bell's theorem~\cite{bell1964einstein} asserts that quantum mechanics cannot
be described by any local hidden-variable theory: measurements performed by
two distant parties on a shared entangled state can produce correlations that
no pair of classical parties, sharing randomness but not
communicating, can reproduce. 
Such nonclassical correlations exhibited by systems operating according to the laws of quantum physics provide the basis for various striking instances of \textit{quantum advantage} in information-processing tasks~\cite{brunner2014bell,buhrman2009non}, and have been confirmed experimentally, see e.g.~the loophole-free Bell tests
of~\cite{hensen2015loophole,giustina2015significant,shalm2015strong}.
In the language of nonlocal games~\cite{cleve2004consequences}, entanglement
allows non-communicating players to coordinate their answers in ways that are
impossible classically, with dramatic computational consequences: for instance, a polynomial-time verifier
interacting with two entangled players can verify membership in any
recursively enumerable language, including the halting problem~\cite{JiNVWY21}.
It is a key challenge, then, to
\emph{quantify} the advantage brought by entanglement in nonlocal games: how much stronger are quantum players than their classical counterparts?

A particularly clean setting in which this question can be addressed
mathematically is the \emph{colouring game}~\cite{galliard2002pseudo}.
Two players, Alice and Bob, are given a graph $G$ and a number $n\in\N$ of
colours. Their goal is to convince a verifier that $G$ admits a proper $n$-colouring---i.e., that it is possible to assign a colour to each vertex of $G$ in a way that no edge is monochromatic. Before the game starts, Alice and Bob may agree on a strategy; they are then separated and can
no longer communicate. The verifier sends a vertex $u$ to Alice and a
vertex $v$ to Bob, and the players answer with colours $a,b\in[n]$. They win
if $a=b$ whenever $u=v$, and $a\neq b$ whenever $u$ and $v$ are adjacent.
Without entanglement, the players can win with certainty---that is, on every
pair of questions---if, and only if, 
the \textit{chromatic number} $\chi(G)$ of $G$ is at most $n$, as a perfect classical
strategy amounts to a proper $n$-colouring of $G$. If, however, Alice and Bob
share an entangled state and answer according to the outcomes of quantum
measurements performed on it, they may be able to win with certainty even if
$\chi(G)>n$. 
This is an example of the phenomenon known as \textit{quantum pseudo-telepathy}~\cite{galliard2002pseudo,brassard2005quantum_pseudo,manvcinska2014maximally}, and it yields an inequality-free proof of Bell’s theorem.
The least $n$ for which this is possible is the \emph{quantum
chromatic number} $\chi_q(G)$~\cite{cameron2006quantum}. Clearly,
$\chi_q(G)\leq\chi(G)$, as entanglement never decreases the power of Alice and Bob. But how much can it \textit{increase} their power? 

\paragraph{The quantum chromatic gap.}
The power of entanglement in the colouring game is measured by the gap between
$\chi_q$ and $\chi$---the \emph{quantum chromatic gap}: is $\chi(G)$ bounded
from above by some function of $\chi_q(G)$? This question is closely related to
the problem of simulating entanglement classically, as it asks whether classical
players can reproduce the behaviour of entangled players if their task is made
easier---here, by giving them more colours.

In fact, the first graphs exhibiting pseudo-telepathy in the colouring game
arose from the study of simulating entanglement with classical communication.
Building on the work of Buhrman, Cleve, and Wigderson~\cite{buhrman1998quantum}
in communication complexity, Brassard, Cleve, and
Tapp~\cite{brassard1999cost} designed a quantum strategy based on the
Deutsch--Jozsa algorithm which, in today's terms, shows that the Hadamard graph
on the set of $n$-bit strings has quantum chromatic number at most $n$ when $n$ is a power of
$2$. On the other hand, a theorem of Frankl and R\"odl~\cite{frankl1987forbidden}
implies that its classical chromatic number is $2^{\Omega(n)}$. Hence,
$\chi$ can be exponential in $\chi_q$.
Pseudo-telepathic colourings and
the quantum chromatic number have since been studied
extensively~\cite{galliard2002pseudo,galliard2002impossibility,avis2006quantum,cameron2006quantum,scarpa2011kochen,paulsen2015quantum,MancinskaR16,paulsen2016estimating,manvcinska2018oddities,elphick2019spectral,lalonde2023quantum,harris2024universality,godsil2024quantum,lalonde2026}.
However, no larger separation than exponential has been found. The current state of the art is due to Lalonde~\cite[Corollary 1.10]{lalonde2026}, who proved that the graph $G_n$ with vertices $\{-1,0,1\}^n\setminus\{(0,\dots,0)\}$ whose edges are given by the orthogonality relation satisfies $\chi(G_n)\geq1.1546^{\chi_q(G_n)}$ for $n$ large enough.

Whether the gap can be unbounded was asked explicitly by Cameron, Montanaro,
Newman, Severini, and Winter~\cite{cameron2006quantum}, by Man\v{c}inska and
Roberson~\cite{MancinskaR16}, and more recently by Lalonde~\cite{lalonde2023quantum}. It was shown in~\cite{cameron2006quantum} that, if the entangled
players may only use measurements consisting of rank-one projectors, then the
chromatic number is at most exponential in the number of colours in a perfect quantum strategy. This suggested that the
exponential gap might be optimal and, in
particular, that the quantum chromatic gap might be bounded---so that, given
sufficiently many colours, classical players could always simulate entangled
ones.

\paragraph{A pseudo-telepathy $d$-to-$1$ conjecture.}
More recently, the first author proposed in~\cite{ciardo2026quantum} that the situation
might be different. The approach of~\cite{ciardo2026quantum} draws on the
theory of hardness of approximation, and specifically on Khot's conjectures on the computational complexity of restrictions of the label-cover problem~\cite{khot2002power}. A \emph{$d$-to-$1$ label-cover game} is a two-prover one-round game in which,
for every pair of questions $(x,y)$ that the verifier may ask, the answer to
$y$ is required to be the image of the answer to $x$ under a fixed map
$\pi_{xy}$ that is $d$-to-$1$. Khot's $d$-to-$1$ conjecture postulates that,
for each $d\geq2$ and each $\epsilon>0$, it is NP-hard to distinguish
$d$-to-$1$ games of value $1$ from $d$-to-$1$ games of value less than
$\epsilon$, for some alphabet size depending only on $d$ and $\epsilon$.
Together with the Unique Games Conjecture, proposed in the same
paper~\cite{khot2002power}, it has been one of the driving forces of the
theory of hardness of approximation, see e.g.~\cite{dinur2006conditional,o2009conditional,guruswami2009improved,khot2014hardness,Khot18:focs-pseudorandom,braverman2019rich,guruswami_et_al:LIPIcs.ICALP.2020.62,braverman2025invariance}.
In a recent breakthrough, Fei, Minzer, and Wang~\cite{fei2026hardness} proved the conjecture
for $d=4$.

The conjecture put forth in~\cite{ciardo2026quantum} is a pseudo-telepathy
analogue of Khot's conjecture: the computational gap between instances of value $1$ and
instances of small value is replaced by a gap between the power of quantum
and classical players in the same protocol. Namely, it postulates the existence of $d$-to-$1$ games that quantum
players can win with certainty using finite-dimensional shared entanglement, but in which classical players cannot win
with probability larger than $\epsilon$---a strong form of pseudo-telepathy.
There is an additional requirement in the  conjecture, having no classical counterpart.
One of the features that most
fundamentally distinguish the quantum world from the classical one is
\emph{contextuality}~\cite{Kochen_specker_hidden,abramsky2011sheaf}: quantum measurements can be
performed jointly only if they are compatible (i.e., if they commute), and
the outcome of a measurement cannot, in general, be regarded as predetermined
independently of its \emph{context}---the set of compatible measurements
performed together with it. In a quantum strategy, a context is thus a set of
questions whose associated measurements pairwise commute. As observed
in~\cite{ciardo2026quantum}, when the reductions from $d$-to-$1$ games to
graph colouring are quantised, the contexts \textit{shrink}: quantum completeness of
the reductions comes at the price of a loss of commutativity. For a quantum strategy to survive the whole chain of
reductions, it must therefore start with large contexts.

The size of the contexts is measured by the notion of compatibility. Roughly
speaking, a \emph{quantum assignment} for a game is a quantum strategy in
which the players perform, depending on their questions, projective
measurements on a shared maximally entangled state; it is \emph{perfect} if
it wins with certainty, and it is \emph{$k$-compatible} if the measurements
associated with any two questions at distance at most $k$ in the constraint
graph (or Gaifman graph) of the game commute; the formal definitions are
recalled in Section~\ref{sec:preliminaries}. For $k=1$, this only asks that
the measurements associated with questions in a common constraint commute; a notion known as \textit{oracularisability}.
As $k$ grows, the measurement contexts grow too, and once $k$ is at least the diameter of
the constraint graph, all measurements commute and the quantum assignment is
essentially classical. 
The conjecture in~\cite{ciardo2026quantum} asks for a strong, highly-compatible form of pseudo-telepathy for $d$-to-$1$ games: there must exist a perfect quantum assignment
that looks classical locally---on any set of questions that are pairwise at
distance at most $k$, it amounts to a probability distribution over classical
assignments satisfying all constraints within the set---and yet, it cannot be turned into any globally classical assignment satisfying a fraction of constraints bounded away from zero.

\begin{conjecture}[{\cite[Conjecture~2]{ciardo2026quantum}}]
\label{conj_d_to_1_khot_quantum}
Fix $d\geq 2$. For each $\epsilon>0$ and each $k\in\N$ there exists a $d$-to-$1$ instance $\Phi$ such that
$\Phi$ admits a perfect $k$-compatible quantum assignment but $\sat(\Phi)<\epsilon$.
\end{conjecture}

The main result of~\cite{ciardo2026quantum} is that
Conjecture~\ref{conj_d_to_1_khot_quantum} would imply that the quantum
chromatic gap is unbounded. Its proof shows that a chain of reductions
transforming $d$-to-$1$ games into instances of approximate graph
colouring---the reductions of Dinur, Mossel, and
Regev~\cite{dinur2006conditional} and of Guruswami and
Sandeep~\cite{guruswami_et_al:LIPIcs.ICALP.2020.62}, followed by multiple rounds of the line-digraph
construction---is \emph{quantum complete}, i.e., that it preserves the
existence of perfect quantum strategies. A key ingredient is a quantum
version of adjunction between Pultr functors, a tool used in the theory of reductions for
promise constraint satisfaction~\cite{barto2021algebraic,krokhin2022invitation,krokhin2023topology}.

\begin{theorem}[{\cite[Theorem~7]{ciardo2026quantum}}]
\label{thm_main_OQCGI}
Suppose that Conjecture~\ref{conj_d_to_1_khot_quantum} holds for some $d\geq 2$. Then, for any $c\in\N$, there exists a graph with quantum chromatic number $3$ and classical chromatic number larger than $c$.
\end{theorem}

\paragraph{Our results.}
In this work, we prove Conjecture~\ref{conj_d_to_1_khot_quantum} for every
$d\geq4$.

\begin{theorem}\label{thm:main-dto1}
Conjecture~\ref{conj_d_to_1_khot_quantum} holds for every $d\geq 4$.
\end{theorem}

Together with Theorem~\ref{thm_main_OQCGI}, this shows, unconditionally,
that the quantum chromatic gap is unbounded.

\begin{theorem}\label{thm:main-graphs}
For every $c\in\N$, there exists a graph with quantum chromatic number $3$
and classical chromatic number larger than $c$.
\end{theorem}

In particular, there is no function $f$ such that
$\chi(G)\leq f(\chi_q(G))$ for every graph $G$: entanglement has
\emph{unbounded} power in the colouring game, and no number of additional
colours allows classical players to simulate entangled players in general.
This improves on the exponential separation
of~\cite{brassard1999cost,lalonde2026}, and shows
that the
restriction to rank-one projectors in the result
of~\cite{cameron2006quantum} is essential. The value $3$ in
Theorem~\ref{thm:main-graphs} is optimal, since a graph with quantum
chromatic number $2$ is bipartite~\cite{cameron2006quantum}.

\paragraph{Proof ideas.}
We now sketch our approach; see Section~\ref{sec_overview} for a more detailed overview.
The starting point of our proof of Theorem~\ref{thm:main-dto1} is the
Mermin--Peres magic square~\cite{mermin1990simple,peres1990incompatible}---a
celebrated proof of the Kochen--Specker theorem~\cite{Kochen_specker_hidden}---which is a system of six
linear equations over $\F$ in nine variables that has no classical solution
but admits a perfect quantum solution.
By viewing it as an equation-vs.-variable game,
the magic square yields a label-cover
instance with a perfect quantum
assignment. The contexts of this assignment, however, are small: a row and a
column of the square that share a variable are at distance $2$ in the
Gaifman graph, and the measurements associated with them do not
commute---this is where contextuality resides.

We massage the magic square in several ways. First, we \emph{stretch} it,
replacing each variable by a long ``wire'' of auxiliary equations that
propagates its value (Section~\ref{sec:magic-square}). This places the six
original equations far apart from each other (see Figure~\ref{fig:stretched}) and makes the quantum
assignment $k$-compatible, for $k$ proportional to the wire length;
the classical value increases, but stays below $1$. Second, we apply
parallel repetition à la~\cite{raz1995parallel}, to make the classical value
arbitrarily small while preserving perfect $k$-compatible quantum assignments
(Section~\ref{sec:repetition}). Third, we pass from label cover to
$3$-uniform hypergraphs through the multi-layered construction of Dinur,
Guruswami, Khot, and Regev~\cite{dinur2003new} and the Kneser-type
hypergraphs of Dinur, Regev, and Smyth~\cite{dinur2005hardness}, in the form
given in~\cite[Appendix~C]{fei2026hardness}, and we show that this
construction is quantum complete and preserves compatibility
(Section~\ref{sec:source}). The outcome is a pseudo-telepathy analogue of
the NP-hardness of colouring $2$-colourable $3$-uniform
hypergraphs~\cite{dinur2005hardness}: for all $c,k\in\N$, we obtain a
$3$-uniform hypergraph with chromatic number larger than $c$ that admits a
perfect $k$-compatible quantum $2$-colouring.

We then feed these hypergraphs into the reduction of Fei, Minzer, and
Wang~\cite{fei2026hardness}, which produces games whose constraints are at
most $4$-to-$1$. We show that this reduction, too, is quantum complete and
preserves compatibility (Section~\ref{sec:fmw}). In our quantum strategy, the
players share one copy of the entangled state of the hypergraph colouring for
each register of the game; they measure, in each register, the colours of
the (at most three) vertices involved in their questions, and answer as the
honest classical provers of~\cite{fei2026hardness} would if these colours
came from a proper $2$-colouring. This follows a scheme similar to the one
used in~\cite{ciardo2026quantum} to establish quantum completeness of the
reduction of Dinur, Khot, Kindler, Minzer, and
Safra~\cite{Khot17:stoc-independent,Khot18:focs-pseudorandom,Dinur18:stoc-non-optimality,Dinur18:stoc-towards},
used to prove NP-hardness of approximating the optimal value of $2$-to-$2$
games with imperfect completeness. Soundness only concerns classical
strategies, so we use the soundness analysis of~\cite{fei2026hardness} as a black
box. Finally, we turn the resulting games into instances that
are \textit{exactly} $d$-to-$1$, for every $d\geq4$, at a small cost in
soundness and preserving perfect $k$-compatible quantum assignments
(Section~\ref{sec:four-to-one}).

Together with the construction in~\cite{ciardo2026quantum},
we thus transfer pseudo-telepathy along the classical chain of reductions that was developed to propagate NP-hardness from
the PCP theorem~\cite{arora1998probabilistic,arora1998proof}, through label
cover and hypergraph colouring~\cite{dinur2005hardness}, to
$d$-to-$1$ games~\cite{fei2026hardness}, and finally to graph colouring~\cite{dinur2006conditional,guruswami_et_al:LIPIcs.ICALP.2020.62}, with the magic
square in place of the PCP theorem. 

\begin{table}[t]
\centering\small
\setlength{\tabcolsep}{4pt}
\renewcommand{\arraystretch}{1.2}
\begin{tabular}{@{}l>{\raggedright\arraybackslash}p{3.9cm}>{\raggedright\arraybackslash}p{3.2cm}>{\raggedright\arraybackslash}p{3.3cm}>{\raggedright\arraybackslash}p{3.7cm}@{}}
\toprule
& \textbf{Transformation} & \textbf{Type of the output} & \textbf{Classical soundness} & \textbf{Quantum completeness}\\
\midrule
(i) & stretching the magic square (Section~\ref{sec:magic-square}) & label cover $\Psi_\lambda$ & the equations sum to $0=1$, so $\sat<1$ & Propositions~\ref{prop:Psi} and~\ref{prop:Psi-compatible}; compatibility $k=4\lambda+1$\\
(ii) & parallel repetition (Section~\ref{sec:repetition}) & label cover $\Psi_\lambda^{\otimes n}$ &
$\sat\leq2^{-\gamma n}$\cite{raz1995parallel} & Proposition~\ref{prop:repetition-assignment}; $k$ preserved\\
(iii) & multi-layered Kneser hypergraph~\cite{dinur2003new,dinur2005hardness} (Section~\ref{sec:source}) & $3$-uniform hypergraph & $\chi>c$; adapted from~\cite[App.~C]{fei2026hardness} & Theorem~\ref{thm:source}; $k$ preserved\\
(iv) & reduction of~\cite{fei2026hardness}, padded to exactly $d$-to-$1$ constraints (Sections~\ref{sec:fmw} and~\ref{sec:four-to-one}) & $d$-to-$1$ instance & $\sat\leq\delta$ \cite{fei2026hardness}; padding adapted from~\cite[App.~B]{fei2026hardness} & Propositions~\ref{prop:fmw-perfect},~\ref{prop:fmw-compatible}, and~\ref{prop:fmw-perfect_new}; $k$ preserved\\
\midrule
(v) & \cite{dinur2006conditional} (see also~\cite{khot2008vertex}) & $d$-to-$d$ instance & small induced-sat value~\cite{dinur2006conditional} & \cite[\S5.1]{ciardo2026quantum}; $k$ halved\\
(vi) & Markov chain on $[2d]^d$~\cite{dinur2006conditional,guruswami_et_al:LIPIcs.ICALP.2020.62} & graph &
no large independent sets~\cite{dinur2006conditional,guruswami_et_al:LIPIcs.ICALP.2020.62} & \cite[\S5.2]{ciardo2026quantum}; $\chi_q\leq2d$, $k$ preserved\\
(vii) & line digraph~\cite{harary1960some}, iterated & graph & $\chi>c$~\cite{HarnerE72,poljak1981arc,poljak1991coloring,zhu1998survey} & \cite[\S\S5.3--5.4]{ciardo2026quantum}; $\chi_q=3$, $k$ roughly halved per iteration\\
\bottomrule
\end{tabular}
\caption{The chain of transformations from the magic square to graphs with
$\chi_q=3$ and $\chi>c$. Steps~(i)--(iv) are the subject of this paper, and
steps~(v)--(vii) are from~\cite{ciardo2026quantum}. The last column indicates
where quantum completeness is proved and how the compatibility $k$ of the
perfect quantum assignment changes.}
\label{tab:pipeline}
\end{table}

\section{Overview of the results and techniques}
\label{sec_overview}
In this section, we give an overview of all steps in our proof of Theorem~\ref{thm:main-dto1}, and we discuss our main techniques. Full details of the proofs, as well as a list of notation and preliminaries, are given in the body of the paper (Sections~\ref{sec:preliminaries}--\ref{sec:four-to-one}).
Table~\ref{tab:pipeline} lists the sequence of transformations that, applied in order to
the Mermin--Peres magic square, produce the graphs of
Theorem~\ref{thm:main-graphs}. Steps~(i)--(iv) are carried out in this paper
and prove Theorem~\ref{thm:main-dto1}, while steps~(v)--(vii) are those
of~\cite{ciardo2026quantum}, which prove Theorem~\ref{thm_main_OQCGI}. Each step
must be \emph{classically sound}---if the input has small classical value (or
large chromatic number), then so does the output---and \emph{quantum
complete}---a perfect $k$-compatible quantum assignment for the input yields a
perfect $k'$-compatible quantum assignment for the output, where $k'$ should not
be much smaller than $k$.

Since soundness is classical, it mostly uses as a black box the soundness
analysis of the corresponding NP-hardness reduction. In particular, we can use the
soundness analyses of~\cite{raz1995parallel} and~\cite{fei2026hardness}, much as integrality gaps of convex relaxations are propagated by
running hardness reductions on integrality-gap instances (see
Section~\ref{sec_conclusion}). The exceptions are step~(iii), 
where we need a slight extension of the analysis in~\cite[Appendix~C]{fei2026hardness} to cover arbitrary biregular
projective label-cover instances; and the simple padding in step~(iv). In
particular, we never need to bound the quantum value of an instance (which will equal
$1$ throughout all steps), so no quantum parallel repetition theorem is needed.

Classically, completeness is the easier direction: honest provers compute their
answers from a global satisfying assignment of the input instance. A perfect
quantum assignment for the input instance provides no such global object, but only measurements, of
which only commuting ones can be performed jointly. Our main task is thus to
show that steps~(i)--(iv) are quantum complete, keeping track of compatibility.
In particular, since steps~(v) and~(vii) shrink the contexts,
Conjecture~\ref{conj_d_to_1_khot_quantum} asks for arbitrarily large
compatibility, and the chain must start from a pseudo-telepathic instance with
a highly compatible perfect quantum assignment. The starting point of
our chain is inherently different from the one in the classical chain of reductions. On the one hand, the PCP theorem is useless here, as its YES instances are
classically satisfiable; on the other hand, the magic square that we use here is an instance of a linear system of equations, whose corresponding perfect-completeness classical gap problem is trivially in P via Gaussian elimination.

\subsection{Quantum assignments and compatibility}\label{subsec:overview-prelims}

We briefly recall the notions needed in this overview; see
Section~\ref{sec:preliminaries} for full details. A CSP instance $\Phi$ consists of a
finite set $X$ of variables, each with a finite alphabet $A_x$, and a
probability distribution $\pi$ over a finite multiset $E$ of constraints, each
consisting of a tuple $\bx=(x_1,\dots,x_r)$ of variables and a predicate
$\phi_\bx\subseteq A_{x_1}\times\dots\times A_{x_r}$; its classical value
$\sat(\Phi)$ is the largest weight of a set of constraints satisfied by an
assignment. If all constraints are binary, we say that $\Phi$ is a label-cover instance; it is bipartite if
$E\subseteq X_1\times X_2$ for some decomposition $X=X_1\sqcup X_2$, projective if each $\phi_\bx$ is the graph of a map
$\pi_\bx\colon A_{x_1}\to A_{x_2}$, and $d$-to-$1$ if, moreover, all fibres of
these maps have size $d$. 

A quantum assignment $Q$ for $\Phi$ over a finite-dimensional Hilbert space
$\cH$ assigns to each variable $x$ a projection-valued measurement (PVM)
$Q(x)=\{Q_{x,a}\}_{a\in A_x}$, i.e., a family of projectors on $\cH$ summing to
the identity. It is perfect if its quantum value $\qsat_Q(\Phi)$ equals $1$, where $\qsat_Q(\Phi)$ is the winning probability of the
strategy in which two players are allowed to measure a shared maximally entangled state
according to $Q$. We say that $Q$ is
\emph{$k$-compatible} if $Q(x)$ and $Q(x')$ commute whenever $x$ and $x'$ have distance at most $k$ in the Gaifman graph of $\Phi$. For $1$-compatible quantum assignments, perfection admits a simple characterisation due to~\cite{abramsky2017quantum}: $Q$
is perfect if, and only if, $Q_{x_1,a_1}\cdots Q_{x_r,a_r}=O$ for every
constraint $\bx$ and every $(a_1,\dots,a_r)\notin\phi_\bx$.

It shall often be useful to modify a quantum assignment by identifying some of the outcomes of the original measurements, exploiting the fact that the class of PVMs is \textit{minor-closed} (i.e., it forms a \textit{minion}, see~\cite{abramsky2017quantum,ciardo2024quantum}).
More precisely, if $\{Q_\omega\}_{\omega\in\Omega}$ is a PVM
and $g\colon\Omega\to B$ is a map, then $Q^g_b=\sum_{\omega:\,g(\omega)=b}Q_\omega$
defines a new PVM $\{Q^g_b\}_{b\in B}$, the \emph{minor} of $\{Q_\omega\}_{\omega\in\Omega}$ along $g$. 
For instance, if $\Phi$ is a projective bipartite label-cover instance and $Q$ is
perfect and $1$-compatible, then $Q(x_2)$ is the minor of $Q(x_1)$ along
$\pi_\bx$ for every constraint $\bx=(x_1,x_2)$
(Lemma~\ref{lem:projective-coarse}).
Minors of
the same PVM commute and multiply like indicator functions of a single random variable:
$Q^g_bQ^h_c=\sum_{\omega:\,g(\omega)=b,\,h(\omega)=c}Q_\omega$
(Lemma~\ref{lem:coarse}).

\subsection{Stretching the magic square}\label{subsec:overview-stretching}

The magic square~\cite{mermin1990simple,peres1990incompatible} is the system $S$ of six linear equations over $\F$ in the nine
variables $x_{rs}$, $r,s\in[3]$, given by
\[
R_r\colon x_{r1}+x_{r2}+x_{r3}=0\quad(r\in[3]),\qquad
C_s\colon x_{1s}+x_{2s}+x_{3s}=b_{C_s}\quad(s\in[3]),
\]
where $b_{C_1}=b_{C_2}=0$ and $b_{C_3}=1$. Since each variable occurs in two
equations, the sum of all equations gives $0=1$, so the system is classically unsatisfiable. In the corresponding
equation-vs.-variable label-cover instance $\Psi(S)$, one player receives an equation and
answers with a satisfying assignment of its three variables, and the other
player receives one of these variables and answers with its value. This
instance admits a perfect $1$-compatible quantum assignment over $\C^4$
(Lemma~\ref{lem:magic}): each variable $x_{rs}$ is assigned a binary PVM
$P_{rs}$, the PVMs of the three variables of each equation commute, and the PVM
of an equation is given by their joint measurement.

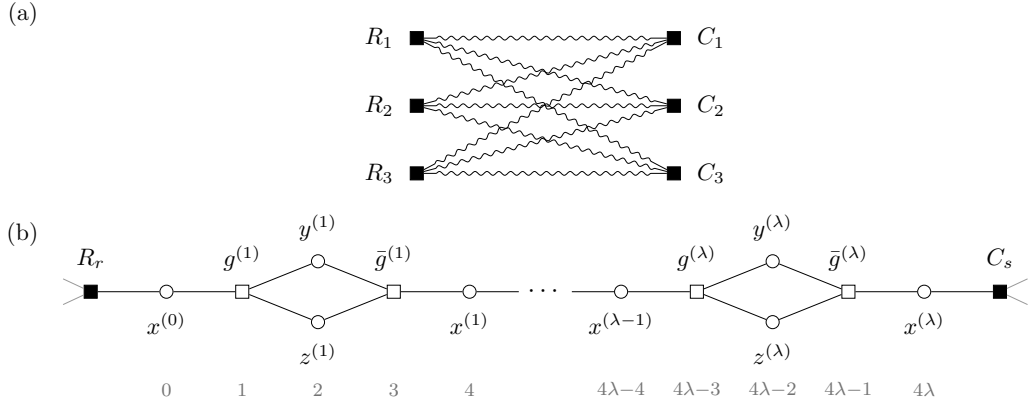
\begin{figure}%
\centering
\scalebox{0.895}{
\begin{tikzpicture}[
orig/.style={draw, rectangle, fill=black, minimum size=5.5pt, inner sep=0pt},
  eq/.style={draw, rectangle, fill=white, minimum size=5.5pt, inner sep=0pt},
  var/.style={draw, circle, fill=white, minimum size=5.5pt, inner sep=0pt},
  wire/.style={decorate, decoration={snake, amplitude=0.8pt, segment length=5pt,               pre length=5pt, post length=5pt}},
stub/.style={black!40},
  lab/.style={font=\small},
  posn/.style={font=\scriptsize, text=black!55},
]
\begin{scope}[xshift=4.82cm]
  \foreach \r in {1,2,3} {
    \node[orig] (R\r) at (0,{-1.0*(\r-1)}) {};
    \node[lab, left=3pt of R\r] {$R_{\r}$};
    \node[orig] (C\r) at (3.8,{-1.0*(\r-1)}) {};
    \node[lab, right=3pt of C\r] {$C_{\r}$};
  }
  \foreach \r in {1,2,3}
    \foreach \s in {1,2,3}
      \draw[wire] (R\r) -- (C\s);
  \node[font=\small] at (-5.82,0.35) {(a)};
\end{scope}
\begin{scope}[yshift=-3.75cm, x=1.12cm]
  \node[font=\small] at (-0.9,0.85) {(b)};
  \node[orig] (R)  at (0,0)     {};
  \node[var]  (x0) at (1,0)     {};
  \node[eq]   (g1) at (2,0)     {};
  \node[var]  (y1) at (3,0.45)  {};
  \node[var]  (z1) at (3,-0.45) {};
  \node[eq]   (h1) at (4,0)     {};
  \node[var]  (x1) at (5,0)     {};
  \node       (dd) at (6,0)     {$\cdots$};
  \node[var]  (xm) at (7,0)     {};
  \node[eq]   (gl) at (8,0)     {};
  \node[var]  (yl) at (9,0.45)  {};
  \node[var]  (zl) at (9,-0.45) {};
  \node[eq]   (hl) at (10,0)    {};
  \node[var]  (xl) at (11,0)    {};
  \node[orig] (C)  at (12,0)    {};
  \draw (R) -- (x0) -- (g1) -- (y1) -- (h1) -- (x1) -- (dd)
        -- (xm) -- (gl) -- (yl) -- (hl) -- (xl) -- (C);
  \draw (g1) -- (z1) -- (h1);
  \draw (gl) -- (zl) -- (hl);
  \draw[stub] (R) -- +(155:0.45cm);
  \draw[stub] (R) -- +(205:0.45cm);
  \draw[stub] (C) -- +(25:0.45cm);
  \draw[stub] (C) -- +(-25:0.45cm);
  \node[lab, above=3pt of R]  {$R_r$};
  \node[lab, above=3pt of C]  {$C_s$};
  \node[lab, below=2pt of x0] {$x^{(0)}$};
  \node[lab, below=2pt of x1] {$x^{(1)}$};
  \node[lab, below=2pt of xm] {$x^{(\lambda-1)}$};
  \node[lab, below=2pt of xl] {$x^{(\lambda)}$};
  \node[lab, above=2pt of g1] {$g^{(1)}$};
  \node[lab, above=2pt of h1] {$\bar g^{(1)}$};
  \node[lab, above=2pt of gl] {$g^{(\lambda)}$};
  \node[lab, above=2pt of hl] {$\bar g^{(\lambda)}$};
  \node[lab, above=2pt of y1] {$y^{(1)}$};
  \node[lab, below=2pt of z1] {$z^{(1)}$};
  \node[lab, above=2pt of yl] {$y^{(\lambda)}$};
  \node[lab, below=2pt of zl] {$z^{(\lambda)}$};
  \foreach \xx/\p in {1/0,2/1,3/2,4/3,5/4,
                      7/4\lambda{-}4,8/4\lambda{-}3,9/4\lambda{-}2,
                      10/4\lambda{-}1,11/4\lambda}
    \node[posn] at (\xx,-1.45) {$\p$};
\end{scope}
\end{tikzpicture}
}
\caption{The stretched system $S_\lambda$ of Definition~\ref{def:stretched}, drawn as the Gaifman graph of
$\Psi(S_\lambda)$. Black squares are the original equations, white squares
are wire equations, circles are variables, and each equation is joined to
the three variables occurring in it. (a) illustrates the original equations; each
wavy line stands for the wire of a pair $(r,s)\in[3]^2$, which joins $R_r$
to $C_s$. (b) illustrates the wire of some fixed $(r,s)$ together with $R_r$ and $C_s$, with the
subscript $rs$ omitted; the grey edges at $R_r$ and at $C_s$ lead to the
other wires. The grey numbers below the wire are the positions of the vertices of the
wire; see Section~\ref{subsec:Psi-compatibility}.}
\label{fig:stretched}
\end{figure}

In order to increase compatibility of the perfect quantum assignment while preserving classical unsatisfiability, we \textit{stretch} $S$: for a stretching parameter $\lambda\in\N$, we split each variable $x_{rs}$ into a copy $x^{(0)}_{rs}$
occurring in $R_r$ and a copy $x^{(\lambda)}_{rs}$ occurring in $C_s$, and we
join the two copies by a \textit{wire}  of $2\lambda$ equations
\[
g^{(i)}_{rs}\colon x^{(i-1)}_{rs}+y^{(i)}_{rs}+z^{(i)}_{rs}=0,\qquad
\bar g^{(i)}_{rs}\colon x^{(i)}_{rs}+y^{(i)}_{rs}+z^{(i)}_{rs}=0\qquad(i\in[\lambda]),
\]
see
Definition~\ref{def:stretched} and Figure~\ref{fig:stretched}. Note that, together, the equations $g^{(i)}_{rs}$ and $\bar g^{(i)}_{rs}$
force $x^{(i-1)}_{rs}=x^{(i)}_{rs}$ and, thus, allow propagating the classical value of the $x$ variables along the entire wire. In the
resulting system $S_\lambda$, every equation still has three variables, every
variable still occurs in two equations, and the sum of all equations
still gives $0=1$. Thus, $\Psi_\lambda=\Psi(S_\lambda)$ is a classically
unsatisfiable, biregular,\footnote{Note: we shall need this in
step~(iii).} projective label-cover instance, all of whose
equations have four satisfying assignments. In Section~\ref{subsec:stretched-assignment}, we construct a
quantum assignment with the following characteristics:
\begin{itemize}[nosep]
\item it keeps the magic-square PVMs on the original equations;
    \item it assigns $P_{rs}$ to the variables $x^{(i)}_{rs}$ and $z^{(i)}_{rs}$, and a
trivial PVM giving all weight to $0$ to the variables $y^{(i)}_{rs}$;
\item and the
wire equations measure $P_{rs}$ and output $(a,0,a)$ upon outcome $a$.
\end{itemize}

\begin{proposition}[Propositions~\ref{prop:Psi} and~\ref{prop:Psi-compatible}]\label{prop:overview-stretching}
For every $\lambda\in\N$, $\sat(\Psi_\lambda)=1-1/(18+54\lambda)$, and
$\Psi_\lambda$ admits a perfect $(4\lambda+1)$-compatible quantum assignment
over $\C^4$.
\end{proposition}

\begin{proof}[Proof sketch]
The classical value for $\sat(\Psi_\lambda)$ is obtained via a simple counting argument.
    The PVM associated with every variable is a minor of the PVM of every equation containing
it; by Lemmas~\ref{lem:perfect} and~\ref{lem:coarse}, this makes the quantum
assignment perfect. For compatibility, observe that all projectors assigned to
the vertices of the wire $\cW_{rs}$ joining $R_r$ and $C_s$ lie in the
commutative algebra $\alg_{rs}$ generated by $P_{rs}$. Likewise, all projectors
assigned to an original equation $u$ lie in the commutative algebra $\alg_u$
generated by the PVMs $P_{rs}$ of the variables $x_{rs}$ of $u$, and
$\alg_{rs}\subseteq\alg_u$ for each such $(r,s)$. Since every wire has length
$4\lambda$, any two original equations are at distance at least $4\lambda+2$.
Hence, a shortest path of length at most $4\lambda+1$ contains at most one
original equation $u$. If it contains none, it lies in a single wire
$\cW_{rs}$; otherwise, it lies in the union of $u$ and of wires attached
to~$u$. In both cases, the projectors assigned to the endpoints of the path lie
in a common commutative algebra ($\alg_{rs}$ or $\alg_u$), so they commute.
\end{proof}

In words, the stretching operation confines contextuality to the six original equations and
moves them \textit{far apart}, so that our quantum assignment behaves classically on
every set of questions of diameter at most $4\lambda+1$. The price is a
classical value very close to $1$, which we address next.

\subsection{Parallel repetition and a pattern for quantum completeness}\label{subsec:overview-pattern}

To decrease the classical value, we apply parallel repetition: by Raz's
theorem~\cite{raz1995parallel}, $\sat(\Psi_\lambda^{\otimes n})\leq2^{-\gamma n}$
for some $\gamma=\gamma(\lambda)>0$ (Theorem~\ref{thm:raz}). 

Quantum completeness is straightforward for parallel repetition (see Proposition~\ref{prop:repetition-assignment}). However, its proof gives a nice instance of a pattern underlying all our
completeness proofs after step~(i); we now describe this pattern at a high level. Suppose that $\Phi'$ is obtained from $\Phi$
by a reduction whose classical completeness is witnessed by honest
provers as follows: the honest answer to a question $z$ of $\Phi'$ is obtained by
applying an encoding $\enc_z$ to the values that a satisfying assignment of
$\Phi$ takes on a few variables, organised in \emph{registers}; each
register $\varrho$ contains a set $F_\varrho(z)$ of variables of $\Phi$, the
\emph{footprint} of $z$. Now, given a perfect $k$-compatible quantum assignment $Q$
for $\Phi$ over $\cH$, the quantum players of $\Phi'$ use one copy of $\cH$ per register.
Upon receiving question $z$, they jointly measure, in each register $\varrho$, the PVMs
$Q(x)$ with $x\in F_\varrho(z)$, and they apply $\enc_z$ to the outcome
$\omega$---their \emph{local view} of a satisfying assignment. In other words,
they measure the minor along $\enc_z$ of the PVM $\Pi_z$ given by
$\Pi_z(\omega)=\bigotimes_\varrho\prod_{x\in F_\varrho(z)}Q_{x,\omega_\varrho(x)}$.
Three properties need to be checked.
\begin{enumerate}[label=(\Alph*),nosep]
\item \emph{Footprints are contexts}: the PVMs $Q(x)$, $x\in F_\varrho(z)$,
  pairwise commute, so that $\Pi_z$ is a PVM. In all our steps, each footprint
  lies within a constraint of $\Phi$, so $1$-compatibility of $Q$ suffices.
\item \emph{Local views are consistent}: for every constraint of $\Phi'$, the
  PVMs $\Pi_z$ of its questions are minors of a single PVM, %
  and every local view
  $\omega$ with nonzero projector is encoded into answers that satisfy the constraint. By Lemmas~\ref{lem:perfect} and~\ref{lem:coarse}, the quantum
  assignment for $\Phi'$ is then perfect.
\item \emph{Footprints move slowly}: if $z$ and $z'$ are adjacent in the Gaifman graph of $\Phi'$, then, in each register $\varrho$, every variable in
  $F_\varrho(z)$ is equal or adjacent to every variable in $F_\varrho(z')$. Then, for any two questions of $\Phi'$ at distance $m\leq k$, the variables in their
  footprints are at distance at most $m$ in each register; by $k$-compatibility
  of $Q$ and Lemma~\ref{lem:tensor-commute}, their PVMs commute, so
  compatibility is preserved.
\end{enumerate}
For parallel repetition, the registers are simply the $n$ coordinates: the players
share $n$ copies of the original entangled state and play each coordinate on its own
copy. The footprint of $\bar x=(\bar x_1,\dots,\bar x_n)$ in coordinate $s$ is
$\{\bar x_s\}$, so a local view is just a tuple of labels
$\bar a=(\bar a_1,\dots,\bar a_n)$, with
$\Pi_{\bar x}(\bar a)=Q_{\bar x_1,\bar a_1}\otimes\dots\otimes Q_{\bar x_n,\bar a_n}$;
the encoding is the identity, i.e., the players answer $\bar a$ itself. This yields a new
quantum assignment $Q^{\otimes n}$. Let us check that it satisfies the three properties. Property~(A) is trivial, as footprints are single
variables. For~(B), if the constraint $(\bar x,\bar x')$ of $\Phi^{\otimes n}$ comes from
the constraints $\bx^{(1)},\dots,\bx^{(n)}$ of $\Phi$, the corresponding PVM in $Q^{\otimes n}$ is the joint measurement of $\bar x$ and $\bar x'$, coordinate by
coordinate, which exists by $1$-compatibility; its outcomes with nonzero projector satisfy
every coordinate as $Q$ is perfect.
For~(C), variables that are adjacent in $\Phi^{\otimes n}$ are adjacent in every
coordinate. Note that a variable of $\Phi$ occurring in two coordinates of $\bar x$ is measured on
two different copies and may receive different labels; this is harmless, as
$\Phi^{\otimes n}$ checks each coordinate separately. For the next reductions, we will
need more creative applications of this pattern: for the Kneser hypergraph, footprints are still singletons and the encoding
records whether a label lies in a given set (Section~\ref{subsec:overview-source});
for the Fei--Minzer--Wang game, footprints lie within hyperedges of the input hypergraph and the encoding
is a quadratic form (Section~\ref{subsec:overview-fmw}). Yet, the pattern \textit{is} the same.

\subsection{The source of pseudo-telepathy}\label{subsec:overview-source}

Next, we turn the instances $\Psi_\lambda^{\otimes n}$ into hypergraphs. Let
$\Phi=(X_1\sqcup X_2,E,A,\pi,\phi)$ be a projective bipartite label-cover
instance, let $c\geq2$, and let $\ell=2c^2$. Following~\cite{dinur2003new}
and~\cite[Appendix~C]{fei2026hardness}, we define the layers
$\cX_i=X_2^{i-1}\times X_1^{\ell-i}$, $i\in[\ell]$. We first construct a graph $\mathcal L$ with vertex set $\bigcup\cX_i$ as follows. For $i<j$, a pair
$(v,v')\in\cX_i\times\cX_j$ is an edge if $v'$ is obtained from $v$ by moving
each coordinate $v_s$ with $i\leq s<j$ along a constraint $(v_s,v'_s)\in E$,
keeping the other coordinates fixed. We assign to $v$ a label set
$A_v=\prod_sA_{v_s}$, and we define the function $\pi_{v,v'}\colon A_v\to A_{v'}$ that applies the maps of
the constraints $(v_s,v'_s)$ to the moved coordinates. The hypergraph
$H_{c,\ell}(\Phi)$, based on the Kneser-type construction
of~\cite{dinur2005hardness}, has a vertex $(v,S)$ for every $v$ and every set
$S\subseteq A_v$ of size $r_v=\lceil(|A_v|-c)/2\rceil$, and a hyperedge
$\{(v,S_1),(v,S_2),(v',S_3)\}$ whenever $(v,v')$ is an edge of $\mathcal L$,
$S_1\cap S_2=\emptyset$, and $\pi_{v,v'}(A_v\setminus(S_1\cup S_2))\subseteq S_3$
(Definition~\ref{def:kneser}). Classically, a satisfying assignment $f$ of
$\Phi$ yields the proper $2$-colouring $(v,S)\mapsto[f(v)\in S]$, where
$f(v)=(f(v_s))_s$.

\begin{proposition}[Proposition~\ref{prop:kneser-colouring} and Theorem~\ref{thm:source-soundness}]\label{prop:overview-source}
Let $\Phi$ be as above, and suppose that $E$ is a set, $\pi$ is uniform, $\Phi$
is biregular, and $|A_v|\geq6c$ for every vertex $v$ of every layer. Let
$N=\max_v|A_v|$.
\begin{enumerate}[label=(\roman*),nosep]
\item If $\Phi$ admits a perfect $k$-compatible quantum assignment, then
  $H_{c,\ell}(\Phi)$ admits a perfect $k$-compatible quantum $2$-colouring.
\item If $\chi(H_{c,\ell}(\Phi))\leq c$, then
  $\sat(\Phi)\geq6^{-c}/\bigl(2c^4(c+1)\ln N\bigr)$.
\end{enumerate}
\end{proposition}

\begin{proof}[Proof sketch]
(i) We follow the pattern of Section~\ref{subsec:overview-pattern}. The
registers are the $\ell-1$ coordinates, and the footprint of a vertex $(v,S)$ in
coordinate $s$ is $\{v_s\}$. Thus, the players measure the label $a$ of $v$
with the PVM $\{W_{v,a}\}_{a\in A_v}$, where
$W_{v,a}=Q_{v_1,a_1}\otimes\dots\otimes Q_{v_{\ell-1},a_{\ell-1}}$, and output
colour $1$ if, and only if, $a\in S$. Given a hyperedge as above, the PVM
$\{W_{v',a'}\}$ is the minor of $\{W_{v,a}\}$ along $\pi_{v,v'}$
(Lemma~\ref{lem:layered}), so the PVMs of the three vertices are minors of
$\{W_{v,a}\}$. Moreover, for every label $a\in A_v$, the colours $[a\in S_1]$,
$[a\in S_2]$, and $[\pi_{v,v'}(a)\in S_3]$ are not all equal: they are not
all $1$ since $S_1\cap S_2=\emptyset$, and not all $0$ by the definition of
hyperedges. This proves property~(B); property~(C) holds because vertices in a
common hyperedge have equal or adjacent first components.

(ii) The argument follows~\cite[Appendix~C]{fei2026hardness}, where it is
carried out for the specific label-cover instances arising in their reduction.
Fix a proper $c$-colouring of $H_{c,\ell}(\Phi)$. For every $v$, a bound on the
chromatic number of Schrijver graphs~\cite{dinur2005hardness} provides a popular
colour $c_v$ whose colour class contains two disjoint sets $S_1(v)$ and
$S_2(v)$; the at most $c$ labels in $B(v)=A_v\setminus(S_1(v)\cup S_2(v))$
serve as candidate labels for $v$. Using that $\Phi$ is biregular, we show that there are layers $i<j$
such that at least a $1/(2c^2)$ fraction of the edges between them have both
endpoints of the same popular colour $c^*$. For such an edge $(v,v')$, the hyperedges
$\{(v,S_1(v)),(v,S_2(v)),(v',S)\}$ prevent every set $S$ of colour $c^*$ from
containing $\pi_{v,v'}(B(v))$, and a counting lemma
of~\cite{fei2026hardness} (Lemma~\ref{lem:fmw-A3}) yields a label of $v'$ that
lies in many of the sets $\pi_{v,v'}(B(v))$. Labelling each $v\in\cX_i$ by a
random element of $B(v)$ then satisfies many of the edges between $\cX_i$ and
$\cX_j$. Finally, each such edge acts on the $i$-th coordinate as a constraint
of $\Phi$, and choosing the other coordinates at random turns this labelling
into an assignment for $\Phi$ of the claimed value.
\end{proof}

Finally, we apply
Proposition~\ref{prop:overview-source} to the instance $\Phi=\Psi_k^{\otimes n}$ from Section~\ref{subsec:overview-pattern}. Its alphabets have size $2^n$ or $4^n$, so
the lower bound in~(ii) decays only like $1/n$, whereas
$\sat(\Psi_k^{\otimes n})\leq2^{-\gamma n}$. Hence, for $n$ large enough,
$\chi(H_{c,\ell}(\Psi_k^{\otimes n}))>c$. On the other hand, by
Proposition~\ref{prop:overview-stretching}, the discussion in Section~\ref{subsec:overview-pattern}, and~(i), this hypergraph admits a
perfect $k$-compatible quantum $2$-colouring over
$(\C^4)^{\otimes n(\ell-1)}$. This yields the following.

\begin{theorem}[Theorem~\ref{thm:source}]
\label{thm_1012_0311}
For all $c,k\in\N$, there exists a $3$-uniform hypergraph $H$ that admits a
perfect $k$-compatible quantum $2$-colouring and satisfies $\chi(H)>c$.
\end{theorem}

\subsection{Quantum completeness of the Fei--Minzer--Wang game}\label{subsec:overview-fmw}

Fei, Minzer, and Wang~\cite{fei2026hardness} reduce colouring $3$-uniform
hypergraphs $H$ to two-prover games $\GameThree[H,\kappa,\ell_1,\ell_2,t,s,\beta]$
whose constraints are at most $4$-to-$1$: if $H$ is $2$-colourable, the game
has value $1$, and if every $4^r$-colouring of $H$ leaves a fraction
$\varepsilon_H$ of monochromatic hyperedges, then, for suitable parameters, the
value of the game is at most $\delta$ (Theorem~\ref{thm:fmw-soundness}). The game is remarkably
complex, but its completeness analysis only involves the following features (see
Section~\ref{subsec:fmw-game}).

\paragraph{Registers and supports.}
Alice's question $q$ contains a sequence $\cE=(E_1,\dots,E_t)$ of \emph{tripled
sets}, where each $E_j\subseteq[\kappa]\times V(H)$ consists of a copy
$\{i\}\times e_{j,i}$ of a hyperedge $e_{j,i}$ of $H$ for each $i\in[\kappa]$.
Bob's question $p$ contains a sequence $\cE'$ with $E'_j=E_j$ for all $j$ but
one, at which some triples are thinned to a single element. Hence, for each
\emph{register} $(j,i)\in\cR=[t]\times[\kappa]$, the support
$\supp_{j,i}(z)=\{v:(i,v)\in Z_j\}$ of a question $z$ with sequence
$\mathcal Z$ (where $\mathcal Z$ is $\cE$ or $\cE'$ depending on who receives the question) is a hyperedge of $H$ if $z$ is asked to Alice, and a hyperedge or
a singleton if $z$ is asked to Bob; moreover,
$\supp_{j,i}(p)\subseteq\supp_{j,i}(q)$ when $(p,q)$ is a constraint
(Lemma~\ref{lem:supports}).

\paragraph{Honest answers.}
A question $z$ also contains a subspace $D_z$ of
$\cV(\mathcal Z)=\bigoplus_j\F^{Z_j}\otimes\F^{Z_j}$, and the players answer
with linear forms on $D_z$. A $2$-colouring $\omega$ of the supports of $z$
determines the linear functionals $h^\omega_j$ on $\F^{Z_j}$ mapping
$e_{(i,v)}$ to the colour of $v$, and the honest answer is
$\enc_z(\omega)=\bigl(\sum_jh^\omega_j\otimes h^\omega_j\bigr)\big|_{D_z}$.
Alice's answer is valid if, for each $j$, it maps to $1$ the vector
\[
n_T=\textstyle\sum_{a=1}^3e_{a}\otimes e_{a}+e_{1}\otimes e_{2}+e_{2}\otimes e_{3}+e_{3}\otimes e_{1}
\]
(where $e_a=e_{(i,v_a)}$)
of every triple $T=\{i\}\times\{v_1,v_2,v_3\}$ of $E_j$, vanishes on all
symmetric tensors $x\otimes y+y\otimes x$, and has rank at most $1$ on a
certain subspace $U_j\otimes U_j$; Bob's answer is valid if it is symmetric of
rank at most $1$ on certain subspaces $U'_j\otimes U'_j$. The verifier accepts if
the two answers agree on a test subspace $L+\bigoplus_jU_j\otimes U_j$ contained in
$D_p\cap D_q$ (Lemma~\ref{lem:fmw-properties}).

\paragraph{The quantum strategy.}
Let $Q$ be a perfect $k$-compatible quantum $2$-colouring of $H$ over $\cH$.
Following Section~\ref{subsec:overview-pattern}, the registers are the elements
of $\cR$, the footprint of $z$ in register $(j,i)$ is $\supp_{j,i}(z)$, and the
encoding is $\enc_z$. The players share a state in $\cH^{\otimes\cR}$; on question $z$,
they measure in each register the colours of the (at most three) vertices of
$\supp_{j,i}(z)$, and answer $\enc_z(\omega)$, where
$\omega=(\omega_{j,i})_{(j,i)\in\cR}$ is the outcome of the measurement. In formulae, the new quantum assignment is then given by
$\hat Q_{z,g}=\sum_{\omega:\,\enc_z(\omega)=g}\Pi_z(\omega)$, where
$\Pi_z(\omega)=\bigotimes_{(j,i)\in\cR}\prod_{v\in\supp_{j,i}(z)}Q_{v,\omega_{j,i}(v)}$.

\begin{theorem}[Propositions~\ref{prop:fmw-perfect}, \ref{prop:fmw-compatible}, and~\ref{prop:fmw-perfect_new}]\label{thm:overview-fmw}
If $H$ admits a perfect $k$-compatible quantum $2$-colouring over $\cH$, then,
for every choice of the parameters, the label-cover instance $\PhiF$ associated
with $\GameThree[H,\kappa,\ell_1,\ell_2,t,s,\beta]$ admits a perfect
$k$-compatible quantum assignment over $\cH^{\otimes\cR}$.
\end{theorem}

\begin{proof}[Proof sketch]
We check the properties of Section~\ref{subsec:overview-pattern}. (A)~holds
because supports are hyperedges or singletons, and $Q$ is $1$-compatible.
(C)~holds because, along a constraint $(p,q)$, the supports in each register lie
within the hyperedge $\supp_{j,i}(q)$. For~(B), let $(p,q)$ be a constraint. As
$\supp_{j,i}(p)\subseteq\supp_{j,i}(q)$, the PVM $\Pi_p$ is the minor of $\Pi_q$
along the restriction map $\omega\mapsto\omega|_p$
(Lemma~\ref{lem:fmw-marginals}$(i)$), so both players' measurements are minors of
$\Pi_q$. Also, if $\Pi_q(\omega)\neq O$, then $\omega$ is not constant on any support
$\supp_{j,i}(q)=\{v_1,v_2,v_3\}$ (Lemma~\ref{lem:fmw-marginals}$(ii)$), since $Q_{v_1,b}Q_{v_2,b}Q_{v_3,b}=O$ by
perfection of $Q$. Writing $b_a=\omega_{j,i}(v_a)$ and using $b_a^2=b_a$, we
obtain
\[
(h^\omega_j\otimes h^\omega_j)(n_T)=b_1+b_2+b_3+b_1b_2+b_2b_3+b_3b_1=1,
\]
as this polynomial evaluates to $1$ precisely when $b_1,b_2,b_3$ are not all
equal. Moreover, $h_j^\omega\otimes h_j^\omega$ vanishes on $x\otimes y+y\otimes x$ and is
symmetric of rank at most $1$. Hence, the encodings $\enc_q(\omega)$ and $\enc_p(\omega|_p)$ of the local view $\omega$
are valid answers, and they agree on the test subspace, since
$h^{\omega|_p}_j$ is a restriction of $h^\omega_j$ (Lemma~\ref{lem:encodings}).
\end{proof}

\paragraph{Soundness.}
We use the soundness analysis of~\cite{fei2026hardness} as a black box
(Theorem~\ref{thm:fmw-soundness}). It requires a robust form of
non-colourability---every $4^r$-colouring of $H$ must leave a fraction
$\varepsilon_H\leq\exp(-\exp(\exp(Cr)))$ of monochromatic hyperedges. For an efficient reduction, $\varepsilon_H$ must not
depend on the size of $H$; for us, it may. If $\chi(H)>4^r$, then every
$4^r$-colouring leaves at least one monochromatic hyperedge, so we can take
$\varepsilon_H=\min\{1/|E(H)|,\exp(-\exp(\exp(Cr)))\}$, at the price of letting the parameters
$\kappa$, $t$, and $\beta$ of the game depend on $|E(H)|$ (Corollary~\ref{cor:fmw-soundness}). Thus,
$\chi(H)>4^r$ suffices, and Theorem~\ref{thm_1012_0311} provides such hypergraphs.

\subsection{The quantum chromatic gap is unbounded}\label{sec:proofs}

The instance $\PhiF$ is at most $4$-to-$1$~\cite[\S6.3]{fei2026hardness}
(Lemma~\ref{lem:at-most-4to1}). To obtain exactly $d$-to-$1$ instances for
every $d\geq4$, we embed the alphabets of the variables in $X_1$ and in $X_2$
into common alphabets of sizes $dN$ and $N$, respectively, and we replace each
constraint---the graph of a partial map with fibres of size at most $d$---by
all its possible extensions to maps with
fibres of size exactly $d$ (Lemma~\ref{lem:completion}). The Gaifman graph does
not change, and the quantum assignment, extended by the zero projector on the
new labels, remains perfect and $k$-compatible; the value of a classical assignment, on the
other hand, increases by at most $d/(dN-M)$ by using new labels, where $M$ is the
largest alphabet size in $X_1$. We now have all ingredients to prove our main
result.

\begin{proof}[Proof of Theorem~\ref{thm:main-dto1}]
Fix $d\geq4$, and choose $\epsilon>0$ and $k\in\N$; we may assume that $\epsilon<1$.
Let $\delta=\epsilon/2$, and choose
$q_0,q_1,q_2,\ell_1,\ell_2,s,r,\vec\tau$ as in
Theorem~\ref{thm:fmw-soundness}. By Theorem~\ref{thm_1012_0311}, applied with
$c=4^r$ and $k$, there is a $3$-uniform hypergraph $H$ with $\chi(H)>4^r$
that admits a perfect $k$-compatible quantum $2$-colouring over some (finite-dimensional)
Hilbert space $\cH$. By Corollary~\ref{cor:fmw-soundness}, there are
$\kappa,t,\beta$ such that $\PhiF=\PhiF(H;\kappa,\ell_1,\ell_2,t,s,\beta)$
satisfies $\sat(\PhiF)\leq\delta$. 
By Lemma~\ref{lem:at-most-4to1}, the instance
$\PhiF$ is at most $4$-to-$1$, hence at most $d$-to-$1$. By Theorem~\ref{thm:overview-fmw}, it admits a
perfect $k$-compatible quantum assignment over $\cH^{\otimes\cR}$. Let $M$
be the largest alphabet size of a Bob question in $\PhiF$, and choose
$N\in\N$ so that $N$ is at least the largest alphabet size of an Alice question, with
$dN>M$ and $d/(dN-M)<\epsilon/2$.
By Lemma~\ref{lem:completion}, there exists a $d$-to-$1$ instance $\Phi=\PhiF^N$ with $\sat(\Phi)<\delta+\epsilon/2=\epsilon$ that
admits a perfect $k$-compatible quantum assignment over $\cH^{\otimes\cR}$.
\end{proof}

Combining Theorem~\ref{thm_main_OQCGI} and Theorem~\ref{thm:main-dto1}, we obtain
Theorem~\ref{thm:main-graphs}.

\begin{remark}\label{rem:k-compatible-colourings}
As observed in~\cite[Remark~32]{ciardo2026quantum}, the proof of~\cite[Theorem~7]{ciardo2026quantum}
can be straightforwardly modified so that, conditionally on Conjecture~\ref{conj_d_to_1_khot_quantum}, for
each $k\in\N$ there exist graphs $G$ of arbitrarily large classical chromatic number
such that $G$ admits a perfect $k$-compatible quantum $3$-colouring. By Theorem~\ref{thm:main-dto1}, this
conclusion holds unconditionally. In particular, for $k=1$, this yields a $3$ vs.\ unbounded gap between $\chi$ and the \textit{oracular} version of the quantum chromatic number, where measurements associated with adjacent vertices are required to commute.
\end{remark}

\begin{remark}\label{rem:dimension}
Fix $c\in\N$, and let $G$ be the graph with $\chi_q(G)=3$ and $\chi(G)>c$ given
by the proof 
of~\cite[Theorem~7]{ciardo2026quantum} for $d=4$, applied to the instances of
Theorem~\ref{thm:main-dto1}. 
All reductions in~\cite{ciardo2026quantum} preserve the Hilbert
space of the perfect quantum assignment~\cite[Remarks~23 and~42]{ciardo2026quantum},
and so does the construction in Lemma~\ref{lem:completion}. Hence, $G$ admits a
perfect $1$-compatible quantum $3$-colouring over the Hilbert space
\[
\bigl((\C^4)^{\otimes n(\ell-1)}\bigr)^{\otimes t\kappa}
\]
of the proof of Theorem~\ref{thm:main-dto1}, whose dimension is
$4^{n(\ell-1)t\kappa}$. Here $r$ is chosen as in Theorem~\ref{thm:fmw-soundness}
with $\delta=\epsilon/2$; $\ell$ and $n$ are the parameters chosen in
Section~\ref{subsec:source-proof} when Theorem~\ref{thm:source} is applied with $4^r$ colours and with
$k$, as in the proof of Theorem~\ref{thm:main-dto1}; and $t$ and $\kappa$ are those of Corollary~\ref{cor:fmw-soundness}.

This dimension necessarily tends to infinity with $c$. Indeed, the quantum
$3$-colouring above is $1$-compatible (the proof of~\cite[Theorem~7]{ciardo2026quantum}
yields $\delta^{(i)}\eta\Phi\rightsquigarrow^1K_3$; see~\cite{ciardo2026quantum} for the terminology), and by~\cite[Theorem~29]{banakh2025classical}
any graph
admitting a perfect $1$-compatible quantum $3$-colouring over a Hilbert space of
dimension $p$ has classical chromatic number at most $3\alpha_p$ for some
constant $\alpha_p$ depending only on
$p$ and satisfying $\alpha_p=(2+o(1))^{p-1}$ as $p\to\infty$.
Furthermore, it was observed in~\cite[Remark~32]{ciardo2026quantum} that the dimension of
any Hilbert space witnessing Conjecture~\ref{conj_d_to_1_khot_quantum} must tend to infinity as
$\epsilon$ tends to~$0$. 
\end{remark}

\section{Conclusion and open problems}
\label{sec_conclusion}
Theorem~\ref{thm:main-graphs} is, ultimately, a statement about the
nonemptiness of a certain class of graphs, defined by properties that, in
principle, have nothing to do with computational efficiency. One would
therefore expect a direct proof, in which neither computability nor
tractability plays any role. Yet, the only route to it that we know of goes
through the machinery of hardness of approximation---label cover, parallel
repetition, multilayered PCPs, and the Grassmann-type encodings
underlying~\cite{fei2026hardness}---which was developed to prove NP-hardness.
At present, complexity-theoretic reductions seem to be our only means of constructing such objects.  We find this situation intriguing.
A similar phenomenon is familiar from the
study of convex relaxations of classical CSPs, where integrality gaps are often obtained by
running hardness reductions on integrality-gap instances of the source
problem; see, e.g.,~\cite{khot2015unique,tulsiani2009csp,Khot18:focs-pseudorandom}.
In the context of $2$-to-$2$
games, this was described as a ``cart before the horse''
phenomenon by Khot, Minzer, and
Safra~\cite{Khot18:focs-pseudorandom}, who pointed out that their
reduction, applied to random instances of $3$-Lin, allows propagating Lasserre
integrality gaps to Unique Games and other graph problems, and no other route appears to be available at present.

We do not give an explicit description of the graphs witnessing
Theorem~\ref{thm:main-graphs}. Nevertheless, they can be reconstructed by
starting from the Mermin--Peres magic square and applying to it, in order,
the following seven transformations as per this paper and~\cite{ciardo2026quantum} (see Table~\ref{tab:pipeline}):
\begin{enumerate}[label=(\roman*),nosep]
\item stretching (Section~\ref{sec:magic-square});
\item parallel repetition (Section~\ref{sec:repetition});
\item the multi-layered Kneser hypergraph (Section~\ref{sec:source});
\item the reduction of~\cite{fei2026hardness} (Sections~\ref{sec:fmw} and~\ref{sec:four-to-one});
\item the chain of transformations of~\cite{dinur2006conditional} (see also~\cite{khot2008vertex}), which
  turns $d$-to-$1$ instances into unweighted $d$-to-$d$ instances, with
  soundness measured by the \textit{induced-sat} value $\operatorname{isat}_t$---the largest
  fraction of variables inducing a sub-instance that can be satisfied by
  assigning at most $t$ labels to each
  variable, see~\cite[\S5.1 and Appendix~A]{ciardo2026quantum};
\item the reduction of~\cite{dinur2006conditional} (for $d=2$) and of~\cite{guruswami_et_al:LIPIcs.ICALP.2020.62} (for general $d$), which uses a
  Markov chain on state set $[2d]^d$ with suitable spectral properties to turn
  $d$-to-$d$ instances into graphs that are quantum $2d$-colourable but
  have no large independent sets~\cite[\S5.2]{ciardo2026quantum};
\item a number of iterations, depending on $d$, of the line-digraph
  construction~\cite{harary1960some}, which
  keeps the classical chromatic
  number large~\cite{HarnerE72,poljak1981arc,poljak1991coloring,zhu1998survey}
  while bringing the number of colours of
  the perfect quantum strategy down to $3$~\cite[\S5.3 and~5.4]{ciardo2026quantum}.
\end{enumerate}
Steps~(ii)--(iv) and~(vi) preserve compatibility, while step~(v) halves
it, and so does, roughly, each iteration in step~(vii); this compatibility
decay is why the large contexts created in step~(i) are needed. Both the
resulting graphs and the dimension of the entangled states shared by the players are enormous. Finding families of small or explicit graphs witnessing the unbounded
$\chi_q$ vs. $\chi$ gap remains an interesting open problem, as does proving or refuting
Conjecture~\ref{conj_d_to_1_khot_quantum} for $d=2,3$.

Finally, we point out that quantum advantage has also been studied in distributed computing, where a
recent line of work is determining for which local graph problems entanglement
helps~\cite{gall2018quantum,coiteux2024no,10.1145/3717823.3718233,balliu2026distributed}.
In the quantum version of the LOCAL model, the nodes of a network may share an
entangled state and exchange quantum messages, and after a bounded number of
communication rounds they must jointly output a labelling of the whole
graph---for instance, a proper colouring. For approximate graph colouring,
entanglement turns out to give no advantage in this
setting~\cite{coiteux2024no}. This does not contradict Theorem~\ref{thm:main-graphs}. A
distributed algorithm must output a single colouring that is valid on every
edge simultaneously, whereas in the colouring game the verifier checks only one
pair of answers at a time. 
Perfect $k$-compatible strategies sit in between, in some sense: on any set of vertices
that are pairwise at distance at most $k$, their measurements commute and yield
a probability distribution over proper colourings of that set, but these local
distributions cannot be glued into a global one. It would be interesting to
understand whether this locally classical behaviour of highly compatible
quantum strategies has a counterpart in distributed models of computation.

\section{Preliminaries}\label{sec:preliminaries}

Throughout this work, $\N=\{1,2,\dots\}$, $[m]=\{1,\dots,m\}$ for $m\in\N$, and $\F$
is the field with two elements. All Hilbert spaces are complex,
finite-dimensional, and nonzero. For a Hilbert space $\cH$, we denote by
$\id_\cH$ the identity and by $O$ the zero map; an (orthogonal) projector on $\cH$ is a linear application $Q$ such that $Q^2=Q=Q^*$;
$[Q,Q']=QQ'-Q'Q$ is the commutator of two
linear applications; and $\tr$ is the (unnormalised) trace. In a graph, a walk of length $m$ is a
sequence of $m+1$ vertices in which consecutive vertices are adjacent; in a
lazy walk, consecutive vertices are allowed to coincide; a path is a walk
whose vertices are pairwise distinct.

\subsection{CSP instances and quantum assignments}\label{subsec:csp}

We define CSPs in the \textit{multisorted} formulation, where distinct variables are assigned potentially different alphabets. This allows writing more compactly the instances arising in some of the intermediate steps of the reduction.

\begin{definition}\label{def:csp}
A \emph{CSP instance} is a tuple $\Phi=(X,E,A,\pi,\phi)$, where
\begin{itemize}[nosep]
\item $X$ is a finite set, whose elements are called variables;
\item $E$ is a finite multiset of constraint tuples $\bx=(x_1,\dots,x_{\ar(\bx)})$
  of variables;
\item $A=(A_x)_{x\in X}$ is a family of finite nonempty sets (alphabets);
\item $\pi$ is a positive probability distribution over $E$;
\item $\phi$ is a collection of predicates
  $\phi_\bx\subseteq A_{x_1}\times\dots\times A_{x_{\ar(\bx)}}$, one for
  each $\bx\in E$.
\end{itemize}
We identify $\phi_\bx$ with its indicator function.
\end{definition}

A \emph{classical assignment} for $\Phi$ is a function $f$ on $X$ with
$f(x)\in A_x$ for every $x\in X$. Its value is
$\sat_f(\Phi)=\mathbb{E}_{\bx\sim\pi}\,\phi_\bx(f(\bx))$, where $f(\bx)$ is
the entrywise application of $f$ to $\bx$. The classical value of $\Phi$ is
$\sat(\Phi)=\max_f\sat_f(\Phi)$.

For a finite set $S$ and a Hilbert space $\cH$, $\PVM_\cH(S)$ is the set of all
\textit{projection-valued measurements} over $\cH$ with outcome set $S$; i.e., it is the set of all
families
$\{Q_s:s\in S\}$ of projectors on $\cH$ with $\sum_{s\in S}Q_s=\id_\cH$. Note that the
members of such a family are pairwise orthogonal. A \emph{quantum assignment over $\cH$} for a CSP instance $\Phi$ assigns to
each variable $x\in X$ a measurement $Q(x)=\{Q_{x,a}:a\in A_x\}\in\PVM_\cH(A_x)$.
Its value is
\[
\qsat_Q(\Phi)=\frac{1}{\dim\cH}\,\mathbb{E}_{\bx\sim\pi}
\sum_{\ba}\phi_\bx(\ba)\,\tr\bigl(Q_{x_1,a_1}\cdots Q_{x_{\ar(\bx)},a_{\ar(\bx)}}\bigr),
\]
where $\ba=(a_1,\dots,a_{\ar(\bx)})$ ranges over
$A_{x_1}\times\dots\times A_{x_{\ar(\bx)}}$; $Q$ is \emph{perfect} if
$\qsat_Q(\Phi)=1$. 

The \emph{Gaifman graph} of $\Phi$ is the undirected graph with vertex set $X$, where
two distinct variables are adjacent if they appear in a common
constraint; by $\dist_\Phi$ we denote the distance in this graph. %
For $k\in\N\cup\{0\}$, a quantum assignment $Q$ for $\Phi$ is
\emph{$k$-compatible} if $[Q_{x,a},Q_{x',a'}]=O$ for all $x,x'\in X$ with
$\dist_\Phi(x,x')\leq k$ and all $a\in A_x$, $a'\in A_{x'}$.
This notion was introduced in~\cite{ciardo2026quantum} in full generality, to capture the 
shrinking of measurement contexts for the quantum strategies induced by reductions encoded by adjunctions of Pultr functors generalising the line-digraph construction (see in particular~\cite[Theorem~5]{ciardo2026quantum}).
For $k=0$, the condition is vacuous, since the members of a PVM commute.
Such quantum assignments, which are sometimes called \emph{non-oracular} in
the literature on nonlocal games, are the ones commonly considered for
games with binary constraints; see, for example,~\cite{MancinskaR16,kempe2010unique}.
For $k=1$, the condition requires the measurements associated with variables
lying in a common constraint to commute. This is typically required for
games with non-binary constraints~\cite{abramsky2017quantum}, and the
corresponding property of quantum strategies is known as
\emph{oracularisability}~\cite{helton2017algebras,JiNVWY21,mousavi2022nonlocal}.
The case $k=2$, in which commutativity is required for measurements
associated with variables at distance at most~$2$, was considered
in~\cite{mousavi2025quantum} to establish quantum completeness of a reduction
between (oracular) Unique Games and MaxCut;
such strategies were called \emph{weak quantum assignments}
therein.

For $1$-compatible quantum assignments, perfection is a condition on products of
projectors.

\begin{lemma}[\protect{\cite{abramsky2017quantum}; see also~\cite[Proposition~12]{ciardo2026quantum}}]\label{lem:perfect}
Let $Q$ be a $1$-compatible quantum assignment for a CSP instance $\Phi$.
Then $Q$ is perfect if, and only if,
$Q_{x_1,a_1}\cdots Q_{x_r,a_r}=O$ for every $\bx=(x_1,\dots,x_r)\in E$ and
every $\ba\notin\phi_\bx$.
\end{lemma}

\paragraph{Label cover.}
A \emph{label-cover instance} is a CSP
instance all of whose constraints are binary; it is \emph{bipartite} if
$E\subseteq X_1\times X_2$ for some decomposition $X=X_1\sqcup X_2$. For $d\in\N$, we say that a
bipartite label-cover instance $\Phi$ is
\emph{at most $d$-to-$1$} if, for every $\bx=(x_1,x_2)\in E$, each
$a_1\in A_{x_1}$ has at most one $a_2\in A_{x_2}$ with
$(a_1,a_2)\in\phi_\bx$, and each $a_2\in A_{x_2}$ has at most $d$ such
$a_1$. Equivalently, $\phi_\bx$ is given by a map
$\pi_\bx\colon D_\bx\to A_{x_2}$ defined on some $D_\bx\subseteq A_{x_1}$
whose fibres have at most $d$ elements. We say that $\Phi$ is
\emph{projective} if, for every $\bx\in E$, $\phi_\bx$ is given by a
map $\pi_\bx\colon A_{x_1}\to A_{x_2}$, i.e., each $a_1\in A_{x_1}$ has
exactly one $a_2\in A_{x_2}$ with $(a_1,a_2)\in\phi_\bx$.\footnote{The maps
$\pi_\bx$, which always carry a subscript, should not be confused with the
distribution $\pi$.}
We say that $\Phi$ is \emph{$d$-to-$1$} if it is projective and, for every
$\bx\in E$, every fibre of $\pi_\bx$ has exactly $d$ elements; in
particular, $|A_{x_1}|=d\,|A_{x_2}|$ for every $(x_1,x_2)\in E$.
We shall need the following simple fact.
\begin{lemma}\label{lem:projective-coarse}
Let $\Phi$ be a projective bipartite label-cover instance, and let $Q$ be a
perfect $1$-compatible quantum assignment for $\Phi$. Then
\[
Q_{x_2,a_2}=\sum_{a_1\in\pi_\bx^{-1}(a_2)}Q_{x_1,a_1}
\qquad\text{for every $\bx=(x_1,x_2)\in E$ and every $a_2\in A_{x_2}$.}
\]
\end{lemma}

\begin{proof}
Since $\Phi$ is projective, $(a_1,b)\notin\phi_\bx$ whenever
$b\neq\pi_\bx(a_1)$; as $Q$ is perfect and $1$-compatible,
Lemma~\ref{lem:perfect} gives $Q_{x_1,a_1}Q_{x_2,b}=O$ for all such pairs.
Hence,
\begin{align*}
Q_{x_2,a_2}
&=\sum_{a_1\in A_{x_1}}Q_{x_1,a_1}Q_{x_2,a_2}
 =\sum_{a_1\in\pi_\bx^{-1}(a_2)}Q_{x_1,a_1}Q_{x_2,a_2}\\
&=\sum_{a_1\in\pi_\bx^{-1}(a_2)}Q_{x_1,a_1}\Bigl(\id-\sum_{b\neq a_2}Q_{x_2,b}\Bigr)
 =\sum_{a_1\in\pi_\bx^{-1}(a_2)}Q_{x_1,a_1}.\qedhere
\end{align*}
\end{proof}

\subsection{Hypergraphs and graphs}\label{subsec:hypergraphs}

A $3$-uniform hypergraph $H$ consists of a finite set $V(H)$ of vertices and
a nonempty set $E(H)$ of $3$-element subsets of $V(H)$, called hyperedges. For
$c\in\N$, a \emph{$c$-colouring} of $H$ is a map $V(H)\to[c]$; it is
\emph{proper} if no hyperedge is monochromatic, and $\chi(H)$ is the least
$c$ for which $H$ has a proper $c$-colouring.

We associate with $H$ the CSP instance
$\Phi_H=(V(H),E_H,A,\pi_H,\phi)$, where
 $E_H$ contains one constraint tuple
$(v_1,v_2,v_3)$ for each hyperedge $\{v_1,v_2,v_3\}\in E(H)$ (with its
vertices listed in some fixed order),
$A_v=\F$ for every $v\in V(H)$,
$\pi_H$ is the uniform distribution
over $E_H$, and $\phi_\bx=\F^3\setminus\{(0,0,0),(1,1,1)\}$ is the Boolean not-all-equal relation for every
$\bx\in E_H$. A \emph{quantum $2$-colouring} of $H$ over a Hilbert space $\cH$ is a
quantum assignment over $\cH$ for $\Phi_H$; it is \emph{perfect},
respectively \emph{$k$-compatible}, if it is so as a quantum assignment for
$\Phi_H$. Two distinct vertices of $H$ are adjacent in the Gaifman graph of
$\Phi_H$ if, and only if, they lie in a common hyperedge; we write $\dist_H$
for $\dist_{\Phi_H}$. By Lemma~\ref{lem:perfect}, a $1$-compatible quantum $2$-colouring $\{Q_{v,b}\}$ of $H$ is perfect if, and only if,
$Q_{v_1,b}Q_{v_2,b}Q_{v_3,b}=O$ for every hyperedge $\{v_1,v_2,v_3\}$ and
every $b\in\F$.

For a graph $G$, $\chi(G)$ denotes its chromatic number, and $\chi_q(G)$ its
\emph{quantum chromatic number}~\cite{cameron2006quantum}: the least $c\in\N$
for which there are a Hilbert space $\cH$ and PVMs
$\{Q_{v,a}\}_{a\in[c]}\in\PVM_\cH([c])$, one for each $v\in V(G)$, such that
$Q_{u,a}Q_{v,a}=O$ for every edge $\{u,v\}$ of $G$ and every $a\in[c]$.

\subsection{Elementary facts about projectors}\label{subsec:coarse}

Given a PVM $\{Q_\omega\}_{\omega\in\Omega}\in\PVM_\cH(\Omega)$ and a map
$g\colon\Omega\to B$ to a finite set $B$, consider the family
$\{Q^g_b\}_{b\in B}$ given by $Q^g_b=\sum_{\omega:\,g(\omega)=b}Q_\omega$.
We observe the following elementary fact.

\begin{lemma}\label{lem:coarse}
Let $\{Q_\omega\}_{\omega\in\Omega}\in\PVM_\cH(\Omega)$, and let
$g\colon\Omega\to B$ and $h\colon\Omega\to C$ be maps to finite sets. 
Then
$\{Q^g_b\}_{b\in B}\in\PVM_\cH(B)$, $\{Q^h_c\}_{c\in C}\in\PVM_\cH(C)$, and
\[
Q^g_bQ^h_c=Q^h_cQ^g_b=\sum_{\omega:\,g(\omega)=b,\ h(\omega)=c}Q_\omega.
\]
\end{lemma}

\begin{proof}
Immediate from $Q_\omega Q_{\omega'}=O$ for $\omega\neq\omega'$ and
$Q_\omega^2=Q_\omega$.
\end{proof}

We call $\{Q^g_b\}_{b\in B}$ the \emph{minor} of $\{Q_\omega\}_{\omega\in\Omega}$
along $g$. This terminology comes from the theory of promise CSPs, see
e.g.~\cite{barto2021algebraic}; in fact, the closure of PVMs under minors corresponds to the fact that PVMs carry the algebraic structure of a \textit{minion}, see~\cite{ciardo2024quantum}.

We shall also repeatedly use the next observation.

\begin{lemma}\label{lem:tensor-commute}
Let $P=P_1\otimes\dots\otimes P_n$ and $P'=P'_1\otimes\dots\otimes P'_n$
be linear applications on $\cH_1\otimes\dots\otimes\cH_n$ such that
$[P_s,P'_s]=O$ for every $s\in[n]$. Then $[P,P']=O$.
\end{lemma}

\begin{proof}
$PP'=\bigotimes_sP_sP'_s=\bigotimes_sP'_sP_s=P'P$.
\end{proof}

\subsection{The magic square}\label{subsec:magic}

We use the notation of~\cite[\S6]{ciardo2026quantum} for systems of linear
equations over $\F$: each equation $u$ is written as
$\sum_{x\in\tilde u}x=b_u$, where $\tilde u$ is the set of variables
appearing in $u$ and $b_u\in\F$. Also, $s(u)$ denotes the set of assignments
$\vartheta\colon\tilde u\to\F$ satisfying $u$. For a
system $T$ of linear equations over $\F$, the
\emph{equation-vs.-variable instance} $\Psi(T)=(X_1\sqcup X_2,E,A,\pi,\phi)$
is the bipartite label-cover instance in which $X_1$ is the set of equations
and $X_2$ the set of variables of $T$; $A_u=s(u)$ for $u\in X_1$ and $A_x=\F$
for $x\in X_2$; $E=\{(u,x):u\in X_1,\ x\in\tilde u\}$; $\pi$ is the uniform
distribution over $E$; and
$\phi_{(u,x)}=\{(\vartheta,\vartheta(x)):\vartheta\in s(u)\}$. Note that
$\Psi(T)$ is projective, with $\pi_{(u,x)}(\vartheta)=\vartheta(x)$.

The Mermin--Peres \textit{magic square}~\cite{mermin1990simple,peres1990incompatible}
is the system $S$ in the nine variables $x_{rs}$, $r,s\in[3]$, consisting of
the six equations
\[
R_r\colon x_{r1}+x_{r2}+x_{r3}=0\quad(r\in[3]),\qquad
C_s\colon x_{1s}+x_{2s}+x_{3s}=b_{C_s}\quad(s\in[3]),
\]
where $b_{C_1}=b_{C_2}=0$ and $b_{C_3}=1$. Each variable $x_{rs}$ occurs in
exactly two equations, namely $R_r$ and $C_s$. Adding all six equations
gives $0=1$, so $S$ has no classical solution. Nevertheless, the following holds.

\begin{lemma}[\cite{mermin1990simple,peres1990incompatible}]\label{lem:magic}
$\Psi(S)$ admits a perfect $1$-compatible quantum assignment over
$\C^4$.
\end{lemma}

We denote this assignment by $\{Q_u\}_{u\in S}\cup\{P_{rs}\}_{r,s\in[3]}$,
where $P_{rs}$ is the PVM assigned to $x_{rs}$. By Lemma~\ref{lem:projective-coarse}, it holds that
\begin{equation}\label{eq:magic-consistency}
\sum_{\substack{\vartheta\in s(u)\\ \vartheta(x_{rs})=b}}Q_{u,\vartheta}=P_{rs,b}
\qquad\text{for all $(r,s)\in[3]^2$, $u\in\{R_r,C_s\}$, and $b\in\F$.}
\end{equation}

\section{Stretching the magic square}\label{sec:magic-square}

Our first step is to modify the instance $\Psi(S)$ of
Section~\ref{subsec:magic} so that its perfect quantum assignment becomes
$k$-compatible for arbitrarily large $k$, while the instance remains
classically unsatisfiable. To this end, we ``stretch'' the magic square:
each variable $x_{rs}$ is split into two copies, one occurring in $R_r$ and
one in $C_s$, which are joined by a long-enough wire of equations forcing
them to be equal. This places any two of the original equations far apart
in the Gaifman graph; see Figure~\ref{fig:stretched}.

\begin{definition}\label{def:stretched}
Fix $\lambda\in\N$.
The system $S_\lambda$ has, for each $(r,s)\in[3]^2$, the variables
$x^{(0)}_{rs},\dots,x^{(\lambda)}_{rs}$ and $y^{(i)}_{rs},z^{(i)}_{rs}$ for
$i\in[\lambda]$, and consists of the following equations:
\begin{itemize}[nosep]
\item the \emph{original equations}
  $R_r\colon x^{(0)}_{r1}+x^{(0)}_{r2}+x^{(0)}_{r3}=0$ for $r\in[3]$, and
  $C_s\colon x^{(\lambda)}_{1s}+x^{(\lambda)}_{2s}+x^{(\lambda)}_{3s}=b_{C_s}$
  for $s\in[3]$, where $b_{C_1}=b_{C_2}=0$ and $b_{C_3}=1$;
\item the \emph{wire equations}
  $g^{(i)}_{rs}\colon x^{(i-1)}_{rs}+y^{(i)}_{rs}+z^{(i)}_{rs}=0$ and
  $\bar g^{(i)}_{rs}\colon x^{(i)}_{rs}+y^{(i)}_{rs}+z^{(i)}_{rs}=0$ for
  $(r,s)\in[3]^2$ and $i\in[\lambda]$.
\end{itemize}
\end{definition}

\begin{lemma}\label{lem:stretched-basic}
The system $S_\lambda$ has $6+18\lambda$ equations and $9+27\lambda$
variables. Each equation involves three distinct variables, each variable
occurs in two equations, and $S_\lambda$ has no classical solution.
\end{lemma}

\begin{proof}
The equation and variable counts are immediate. The variable $x^{(0)}_{rs}$ occurs in $R_r$ and
$g^{(1)}_{rs}$; $x^{(\lambda)}_{rs}$ occurs in $C_s$ and
$\bar g^{(\lambda)}_{rs}$; for $0<i<\lambda$, $x^{(i)}_{rs}$ occurs in
$\bar g^{(i)}_{rs}$ and $g^{(i+1)}_{rs}$; and $y^{(i)}_{rs}$ and
$z^{(i)}_{rs}$ occur in $g^{(i)}_{rs}$ and $\bar g^{(i)}_{rs}$; see also Figure~\ref{fig:stretched}. Adding all
equations, every variable cancels, while the right-hand sides add up to
$b_{C_3}=1$.
\end{proof}

\subsection{The quantum assignment}\label{subsec:stretched-assignment}

\begin{proposition}\label{prop:stretched-assignment}
There are PVMs $Q_u\in\PVM_{\C^4}(s(u))$, one for each equation
$u$ of $S_\lambda$, and $Q_x\in\PVM_{\C^4}(\F)$, one for each
variable $x$ of $S_\lambda$, such that
\begin{equation}\label{eq:marginal}
Q_{x,b}=\sum_{\substack{\vartheta\in s(u)\\ \vartheta(x)=b}}Q_{u,\vartheta}
\qquad\text{for every equation $u$, every $x\in\tilde u$, and every $b\in\F$.}
\end{equation}
Specifically, for all $(r,s)\in[3]^2$:
\begin{itemize}[nosep]
\item $Q_{x^{(i)}_{rs}}=P_{rs}$ for $0\leq i\leq\lambda$, and, for $i\in[\lambda]$,
  $Q_{z^{(i)}_{rs}}=P_{rs}$, $Q_{y^{(i)}_{rs},0}=\id$, and $Q_{y^{(i)}_{rs},1}=O$;
\item $Q_{R_r}$ and $Q_{C_s}$ are the PVMs of Lemma~\ref{lem:magic};
\item for $i\in[\lambda]$ and $u=g^{(i)}_{rs}$, the satisfying assignments are the triples
  $(a,c,a+c)$ of values on $(x^{(i-1)}_{rs},y^{(i)}_{rs},z^{(i)}_{rs})$,
  and we let $Q_{u,(a,0,a)}=P_{rs,a}$ and $Q_{u,(a,1,a+1)}=O$; likewise
  for $u=\bar g^{(i)}_{rs}$, with $x^{(i)}_{rs}$ in place of
  $x^{(i-1)}_{rs}$.
\end{itemize}
\end{proposition}

\begin{proof}
It is straightforward to check that all the families listed are PVMs. For an original
equation $u$, \eqref{eq:marginal} is~\eqref{eq:magic-consistency}. For $u=g^{(i)}_{rs}$, the right-hand side
of~\eqref{eq:marginal} equals
$Q_{u,(b,0,b)}+Q_{u,(b,1,b+1)}=P_{rs,b}$ for $x=x^{(i-1)}_{rs}$; it equals
$Q_{u,(0,0,0)}+Q_{u,(1,0,1)}=P_{rs,0}+P_{rs,1}=\id$ for $x=y^{(i)}_{rs}$ and $b=0$; it equals
$Q_{u,(0,1,1)}+Q_{u,(1,1,0)}=O+O=O$ for $x=y^{(i)}_{rs}$ and $b=1$; and it equals
$Q_{u,(b,0,b)}+Q_{u,(b+1,1,b)}=P_{rs,b}$ for $x=z^{(i)}_{rs}$. 
In all cases, this coincides with the left-hand side of~\eqref{eq:marginal}.
The case
$u=\bar g^{(i)}_{rs}$ is identical.
\end{proof}

\subsection{The equation-vs.-variable instance}\label{subsec:Psi}

For $\lambda\in\N$, let $\Psi_\lambda=\Psi(S_\lambda)$; i.e., $\Psi_\lambda=(X_1\sqcup X_2,E,A,\pi,\phi)$ is the bipartite
label-cover instance in which $X_1$ is the set of equations of the system $S_\lambda$ and $X_2$ is the
set of variables of $S_\lambda$; $A_u=s(u)$ for $u\in X_1$ and $A_x=\F$ for
$x\in X_2$; $E=\{(u,x):u\in X_1,\ x\in\tilde u\}$; $\pi$ is the uniform
distribution over $E$; and $\phi_{(u,x)}=\{(\vartheta,\vartheta(x)):\vartheta\in s(u)\}$.

\begin{proposition}\label{prop:Psi}
\begin{enumerate}[label=(\roman*)]
\item $\Psi_\lambda$ is projective, $E$ is a set, and for every
  $(u,x)\in E$ the map $\vartheta\mapsto\vartheta(x)$ from $s(u)$ to $\F$ is
  exactly $2$-to-$1$. Every $u\in X_1$ lies in exactly $3$ constraints and
  every $x\in X_2$ in exactly $2$.
\item The PVMs of Proposition~\ref{prop:stretched-assignment} form a perfect
  quantum assignment for $\Psi_\lambda$ over $\C^4$.
\item $\sat(\Psi_\lambda)=1-1/(18+54\lambda)$.
\end{enumerate}
\end{proposition}

\begin{proof}
(i) Each $u$ involves three distinct variables, $s(u)$ consists of four
assignments, and for each $x\in\tilde u$ and $b\in\F$ exactly two of them
satisfy $\vartheta(x)=b$. The degrees follow from
Lemma~\ref{lem:stretched-basic}.

(ii) The Gaifman graph of $\Psi_\lambda$ is bipartite, with $u$ adjacent to
$x$ if and only if $x\in\tilde u$. For such a pair, \eqref{eq:marginal} and
Lemma~\ref{lem:coarse} show that $Q_{u,\vartheta}$ commutes with $Q_{x,b}$,
and that $Q_{u,\vartheta}Q_{x,b}=O$ unless $\vartheta(x)=b$. Thus the
assignment is $1$-compatible, and it is perfect by Lemma~\ref{lem:perfect}.

(iii) Given a classical assignment $f$, let $g=f|_{X_2}$, viewed as an
assignment to the variables of $S_\lambda$. A constraint $(u,x)$ is
satisfied only if $f(u)(x)=g(x)$. If $u$ is violated by $g$, then
$f(u)\neq g|_{\tilde u}$, so at least one of the three constraints at $u$ is
violated. 
By Lemma~\ref{lem:stretched-basic}, $g$ violates at least one
equation, so $\sat_f(\Psi_\lambda)\leq1-1/|E|$ with
$|E|=3(6+18\lambda)=18+54\lambda$. Conversely, let $g$ be the all-zero
assignment. The only equation violated by $g$ is $C_3$, whose right-hand side is $1$. Letting $f(u)=g|_{\tilde u}$ for $u\neq C_3$ and
$f(C_3)$ be the assignment with values $(0,0,1)$ on
$(x^{(\lambda)}_{13},x^{(\lambda)}_{23},x^{(\lambda)}_{33})$, exactly one
constraint, $(C_3,x^{(\lambda)}_{33})$, is violated.
\end{proof}

\subsection{Compatibility}\label{subsec:Psi-compatibility}

For $(r,s)\in[3]^2$, the \emph{wire} $\cW_{rs}$ is the set of the $5\lambda+1$
vertices $x^{(i)}_{rs}$ ($0\leq i\leq\lambda$), $y^{(i)}_{rs}$,
$z^{(i)}_{rs}$, $g^{(i)}_{rs}$, $\bar g^{(i)}_{rs}$ ($i\in[\lambda]$) of the
Gaifman graph of $\Psi_\lambda$.
Following Figure~\ref{fig:stretched}(b), define the position of these vertices by
\[
\operatorname{pos}(x^{(i)}_{rs})=4i,\quad
\operatorname{pos}(g^{(i)}_{rs})=4i-3,\quad
\operatorname{pos}(y^{(i)}_{rs})=\operatorname{pos}(z^{(i)}_{rs})=4i-2,\quad
\operatorname{pos}(\bar g^{(i)}_{rs})=4i-1.
\]

The next lemma collects some properties of wires (which can be easily derived from Figure~\ref{fig:stretched}).

\begin{lemma}\label{lem:wires}
\begin{enumerate}[label=(\roman*)]
\item The six original equations and the nine wires partition the vertex
  set of the Gaifman graph of $\Psi_\lambda$.
\item The only edges between $\cW_{rs}$ and the rest of the graph are
  $\{R_r,x^{(0)}_{rs}\}$ and $\{C_s,x^{(\lambda)}_{rs}\}$.
\item Adjacent vertices of $\cW_{rs}$ have positions differing by exactly
  $1$.
\item Every walk between two distinct original equations has length at
  least $4\lambda+2$.
\end{enumerate}
\end{lemma}

\begin{proof}
Parts (i)--(iii) follow directly from Definition~\ref{def:stretched}.
For
(iv), let $w_0,\dots,w_m$ be a walk with $w_0$ and $w_m$ distinct original
equations. Let $b$ be the least index such that $w_b$ is an original
equation different from $w_0$, and let $a<b$ be the largest index such that
$w_a$ is an original equation; by the choice of $b$, $w_a=w_0\neq w_b$.
Since equations are adjacent only to variables, $b\geq a+2$. The vertices
$w_{a+1},\dots,w_{b-1}$ are not original equations, so by (ii) they
lie in a single wire $\cW_{rs}$, and $w_{a+1}$ and $w_{b-1}$ are adjacent to
$w_a$ and $w_b$, respectively. By (ii), $\{w_a,w_b\}=\{R_r,C_s\}$ and
$\{w_{a+1},w_{b-1}\}=\{x^{(0)}_{rs},x^{(\lambda)}_{rs}\}$, whose positions
are $0$ and $4\lambda$. By (iii), $b-a-2\geq4\lambda$, so $m\geq 4\lambda+2$.
\end{proof}

\begin{proposition}\label{prop:Psi-compatible}
The perfect quantum assignment for $\Psi_\lambda$ of Proposition~\ref{prop:Psi}(ii) is
$k$-compatible for every $k\leq4\lambda+1$. %
\end{proposition}

\begin{proof}
For $(r,s)\in[3]^2$, let $\alg_{rs}$ be the algebra generated by $\id$,
$P_{rs,0}$ and $P_{rs,1}$; note that it is commutative. For an original equation $u$,
let $\alg_u$ be the algebra generated by $\id$ and the projectors $Q_{x,b}$
with $x\in\tilde u$ and $b\in\F$. By~\eqref{eq:marginal} and
Lemma~\ref{lem:coarse}, these projectors belong to minors of $Q_u$, so
$\alg_u$ is commutative, too, and $\alg_u$ contains
$Q_{u,\vartheta}=\prod_{x\in\tilde u}Q_{x,\vartheta(x)}$ for every
$\vartheta\in s(u)$. Moreover, $\alg_{rs}\subseteq\alg_{R_r}\cap\alg_{C_s}$,
since $Q_{x^{(0)}_{rs}}=Q_{x^{(\lambda)}_{rs}}=P_{rs}$. By
Proposition~\ref{prop:stretched-assignment}, every projector assigned to a
vertex of $\cW_{rs}$ lies in $\alg_{rs}$.

Let $w\neq w'$ be vertices of the Gaifman graph of $\Psi_\lambda$ with $\dist_{\Psi_\lambda}(w,w')\leq k\leq4\lambda+1$,
and fix a shortest path between them. By Lemma~\ref{lem:wires}(iv), this path
contains at most one original equation. If it contains none, all its
vertices lie in a single wire $\cW_{rs}$ by Lemma~\ref{lem:wires}(i)--(ii), so
the projectors assigned to $w$ and $w'$ lie in the commutative algebra
$\alg_{rs}$. If it contains exactly one original equation $u$, then removing $u$ splits
the path into at most two subpaths containing no original equation. By
Lemma~\ref{lem:wires}(i)--(ii), each of them lies in a single wire, and
since it contains a neighbour of $u$, by Lemma~\ref{lem:wires}(ii) this wire
must be $\cW_{rs}$ for some $(r,s)$ with $u\in\{R_r,C_s\}$; i.e., $x_{rs}$
must occur in the equation $u$ of the system $S$. For each such $(r,s)$,
$\alg_{rs}\subseteq\alg_u$; hence, the projectors assigned to $w$ and $w'$
lie in the commutative algebra $\alg_u$.
\end{proof}

In particular, the result above implies that, for every
$k\in\N$, the instance $\Psi_k$ admits a perfect $k$-compatible quantum
assignment over $\C^4$. 

\section{Parallel repetition}\label{sec:repetition}

The stretched magic square $\Psi_\lambda$ of Section~\ref{sec:magic-square} admits
perfect, highly compatible quantum assignments, and it is classically unsatisfiable; however, its classical value is close to $1$. Before
feeding it into the multi-layered construction of Dinur, Guruswami, Khot,
and Regev~\cite{dinur2003new} in Section~\ref{sec:source}, we reduce its
classical value by parallel repetition.

\begin{definition}\label{def:repetition}
Let $\Phi=(X_1\sqcup X_2,E,A,\pi,\phi)$ be a bipartite label-cover instance,
and take $n\in\N$. The \emph{$n$-fold repetition} $\Phi^{\otimes n}$ of $\Phi$ is the
bipartite label-cover instance with variable set $X_1^n\sqcup X_2^n$ and
alphabets
$A_{\bar x}=A_{\bar x_1}\times\dots\times A_{\bar x_n}$ for
$\bar x=(\bar x_1,\dots,\bar x_n)$; for each $n$-tuple
$(\bx^{(1)},\dots,\bx^{(n)})$ of constraints of $\Phi$ (taken with
multiplicities), it has the constraint tuple
$\bigl((x^{(1)}_1,\dots,x^{(n)}_1),(x^{(1)}_2,\dots,x^{(n)}_2)\bigr)$, of
weight $\prod_s\pi(\bx^{(s)})$ and predicate
$\{(\bar a,\bar a'):(\bar a_s,\bar a'_s)\in\phi_{\bx^{(s)}}\text{ for all }s\in[n]\}$.
\end{definition}

If $E$ is a set, then so is the multiset of constraint tuples of
$\Phi^{\otimes n}$, since the pairs $\bx^{(s)}$ can be recovered from the
tuples of variables. If $\Phi$ is projective, so is $\Phi^{\otimes n}$, with
maps $\bar a\mapsto(\pi_{\bx^{(s)}}(\bar a_s))_{s\in[n]}$. If $\pi$ is
uniform and every variable in $X_1$ (respectively, $X_2$) lies in exactly
$d_1$ (respectively, $d_2$) constraints, then the weights of
$\Phi^{\otimes n}$ are uniform and every variable in $X_1^n$ (respectively,
$X_2^n$) lies in exactly $d_1^n$ (respectively, $d_2^n$) constraints.

When $E$ is a set, $\Phi$ is a two-prover one-round game of value
$\sat(\Phi)$, in which the verifier samples $\bx=(x_1,x_2)\sim\pi$, sends $x_1$ to the first prover and $x_2$ to the second, and
accepts answers $(a_1,a_2)$ if, and only if, $(a_1,a_2)\in\phi_\bx$. Hence, Raz's
theorem~\cite{raz1995parallel} yields the following.

\begin{theorem}[{\cite{raz1995parallel}}]\label{thm:raz}
Let $\Phi$ be a bipartite label-cover instance in which $E$ is a set, and
suppose that $\sat(\Phi)<1$. Then there is $\gamma_\Phi>0$ such that
$\sat(\Phi^{\otimes n})\leq2^{-\gamma_\Phi n}$ for every $n\in\N$.
\end{theorem}

The next, straightforward argument shows that parallel repetition
preserves perfect $k$-compatible quantum assignments, so that it can be safely applied to the instance of Section~\ref{sec:magic-square}.

\begin{proposition}\label{prop:repetition-assignment}
Let $\Phi$ be a bipartite label-cover instance, and let $Q$ be a
perfect $k$-compatible quantum assignment for $\Phi$ over some Hilbert space $\cH$, with
$k\in\N$.
Then
\[
Q^{\otimes n}_{\bar x,\bar a}:=Q_{\bar x_1,\bar a_1}\otimes\dots\otimes Q_{\bar x_n,\bar a_n}
\]
defines a perfect $k$-compatible quantum assignment for $\Phi^{\otimes n}$
over $\cH^{\otimes n}$. 
\end{proposition}

\begin{proof}
Clearly, $Q^{\otimes n}(\bar x)\in\PVM_{\cH^{\otimes n}}(A_{\bar x})$ for each variable $\bar x$. If
$\bar x$ and $\bar x'$ are adjacent in the Gaifman graph of
$\Phi^{\otimes n}$, then $\bar x_s$ and $\bar x'_s$ are adjacent in the
Gaifman graph of $\Phi$ for every $s$; hence,
$\dist_\Phi(\bar x_s,\bar x'_s)\leq\dist_{\Phi^{\otimes n}}(\bar x,\bar x')$,
and $k$-compatibility follows from Lemma~\ref{lem:tensor-commute}. 
For a constraint $(\bar x,\bar x')$ of $\Phi^{\otimes n}$ coming from
$(\bx^{(1)},\dots,\bx^{(n)})$ and a pair $(\bar a,\bar a')$ outside its
predicate, some coordinate $s$ satisfies
$(\bar a_s,\bar a'_s)\notin\phi_{\bx^{(s)}}$; by Lemma~\ref{lem:perfect}
applied to $\Phi$, the corresponding tensor factor of
$Q^{\otimes n}_{\bar x,\bar a}Q^{\otimes n}_{\bar x',\bar a'}$ vanishes. Thus, the quantum assignment
$Q^{\otimes n}$ is perfect by Lemma~\ref{lem:perfect}.
\end{proof}

\section{The source of pseudo-telepathy}\label{sec:source}

In this section, we construct a strongly pseudo-telepathic $3$-uniform hypergraph that we shall later feed into the Fei--Minzer--Wang reduction in order to prove Conjecture~\ref{conj_d_to_1_khot_quantum}. Concretely, we shall prove the following.

\begin{theorem}\label{thm:source}
For all $c,k\in\N$, there exists a $3$-uniform hypergraph $H$ that admits a
perfect $k$-compatible quantum $2$-colouring and satisfies $\chi(H)>c$.
\end{theorem}
To prove this result, we make use of the
multi-layered PCP of Dinur, Guruswami, Khot, and
Regev~\cite{dinur2003new} and the hypergraph construction of Dinur, Regev, and
Smyth~\cite{dinur2005hardness}, in the form of~\cite[Appendix~C]{fei2026hardness}, applied to the stretched magic square instance of
Sections~\ref{sec:magic-square} and~\ref{sec:repetition}.

\subsection{A multi-layered PCP construction}\label{subsec:layers}

Throughout Sections~\ref{subsec:layers} and~\ref{subsec:kneser},
$\Phi=(X_1\sqcup X_2,E,A,\pi,\phi)$ is a projective bipartite label-cover
instance in which $E$ is a set; for a constraint $\bx=(x_1,x_2)\in E$, we write
$\pi_{x_1x_2}\colon A_{x_1}\to A_{x_2}$ for the map whose graph is
$\phi_\bx$. Moreover, $Q$ is a perfect $k$-compatible quantum assignment for
$\Phi$ over a Hilbert space $\cH$, with $k\in\N$.

Fix $\ell\geq2$. For $i\in[\ell]$, following the presentation of~\cite[Appendix~C, Step~1]{fei2026hardness}, define the $i$-th layer
$\cX_i=X_2^{i-1}\times X_1^{\ell-i}$; different layers are disjoint, since
$X_1\cap X_2=\emptyset$. For
$v=(v_1,\dots,v_{\ell-1})\in\cX_i$, let
$A_v=A_{v_1}\times\dots\times A_{v_{\ell-1}}$. For $1\leq i<j\leq\ell$, let
$\cE_{i,j}$ be the set of pairs $(v,v')\in\cX_i\times\cX_j$ such that
$v_s=v'_s$ for $s\notin\{i,\dots,j-1\}$ and $(v_s,v'_s)\in E$ for
$s\in\{i,\dots,j-1\}$. For such a pair, let $\pi_{v,v'}\colon A_v\to A_{v'}$
act as $\pi_{v_sv'_s}$ on the coordinates $s\in\{i,\dots,j-1\}$ and as the
identity on the other coordinates.  Let $\cL$ be the graph on vertex set
$\bigcup_i\cX_i$ whose edges are the pairs $\{v,v'\}$ with
$(v,v')\in\cE_{i,j}$ for some $i<j$. For $v\in\bigcup_i\cX_i$ and $a\in A_v$,
let
\begin{equation}
W_{v,a}=Q_{v_1,a_1}\otimes\dots\otimes Q_{v_{\ell-1},a_{\ell-1}}\label{W-from-Q}
\end{equation}
be the corresponding projector on $\cH^{\otimes(\ell-1)}$; clearly,
$\{W_{v,a}:a\in A_v\}\in\PVM_{\cH^{\otimes(\ell-1)}}(A_v)$.

\begin{lemma}\label{lem:layered}
\begin{enumerate}[label=(\roman*)]
\item For $(v,v')\in\cE_{i,j}$ and $a'\in A_{v'}$,
  $W_{v',a'}=\sum_{a\in\pi_{v,v'}^{-1}(a')}W_{v,a}$.
\item If $\dist_\cL(v,v')\leq k$, then $[W_{v,a},W_{v',a'}]=O$ for all $a\in A_v$ and $a'\in A_{v'}$.
\end{enumerate}
\end{lemma}

\begin{proof}
(i) Immediate from Lemma~\ref{lem:projective-coarse}, since
$\pi_{v,v'}^{-1}(a')$ is the product of the sets $\pi_{v_sv'_s}^{-1}(a'_s)$
for $s\in\{i,\dots,j-1\}$ and of the singletons $\{a'_s\}$ for the other
coordinates. (ii) Along an edge of $\cL$, each coordinate either stays fixed
or moves along an edge of the Gaifman graph of $\Phi$; hence, for every
$s$, the $s$-th coordinates of $v$ and $v'$ are joined by a lazy walk of
length at most $k$ in that graph, and so are at distance at most $k$. The
claim then follows from $k$-compatibility of $Q$ and
Lemma~\ref{lem:tensor-commute}.
\end{proof}

\subsection{The Kneser hypergraph and its quantum colouring}\label{subsec:kneser}

\begin{definition}[\protect{\cite{dinur2005hardness},\cite[Appendix~C, Step~2]{fei2026hardness}}]\label{def:kneser}
Let $c\geq2$, and suppose that $|A_v|\geq c+1$ for every
$v\in\bigcup_i\cX_i$. For such $v$, let $r_v=\lceil(|A_v|-c)/2\rceil$. The
$3$-uniform hypergraph $H_{c,\ell}(\Phi)$ has vertex set
$\{(v,S):v\in\bigcup_i\cX_i,\ S\in\binom{A_v}{r_v}\}$, and its hyperedges
are the sets $\{(v,S_1),(v,S_2),(v',S_3)\}$ such that $(v,v')\in\cE_{i,j}$
for some $i<j$, $S_1\cap S_2=\emptyset$, and
$\pi_{v,v'}\bigl(A_v\setminus(S_1\cup S_2)\bigr)\subseteq S_3$.
\end{definition}

Note that $r_v\geq1$ and $c-1\leq|A_v|-2r_v\leq c$. The three vertices of a
hyperedge are distinct, since $S_1\cap S_2=\emptyset$ and $v\neq v'$ lie in
different layers.

\begin{proposition}\label{prop:kneser-colouring}
Let $Q$ be a perfect $k$-compatible quantum assignment for $\Phi$ over $\cH$, and let $W=\{W_{v,a}\}$ be defined as in (\ref{W-from-Q}).
For every vertex $(v,S)$ of $H_{c,\ell}(\Phi)$, let
\[
Q'_{(v,S),1}=\sum_{a\in S}W_{v,a},\qquad Q'_{(v,S),0}=\id-Q'_{(v,S),1}.
\]
This defines a perfect $k$-compatible quantum $2$-colouring of
$H_{c,\ell}(\Phi)$ over $\cH^{\otimes(\ell-1)}$.
\end{proposition}

\begin{proof}
The family $\{Q'_{(v,S),0},Q'_{(v,S),1}\}$ is a minor of $W_v$, hence it is a PVM by Lemma~\ref{lem:coarse}. If two
vertices $(v,S)$ and $(v',S')$ of $H_{c,\ell}(\Phi)$ lie in a common
hyperedge, then either $v=v'$ or $\{v,v'\}$ is an edge of $\cL$. Hence a
walk of length $m$ in the Gaifman graph of $H_{c,\ell}(\Phi)$ from $(v,S)$ to $(v',S')$ gives
a lazy walk of length $m$ in $\cL$ from $v$ to $v'$, and $k$-compatibility of $Q'$ follows from
Lemma~\ref{lem:layered}(ii) and linearity of the commutator.

Consider now a hyperedge $\{(v,S_1),(v,S_2),(v',S_3)\}$ with
$(v,v')\in\cE_{i,j}$. As $Q'$ is $1$-compatible, it suffices by
Lemma~\ref{lem:perfect} to show that
$Q'_{(v,S_1),b}Q'_{(v,S_2),b}Q'_{(v',S_3),b}=O$ for $b\in\F$. By
Lemma~\ref{lem:layered}(i),
$Q'_{(v',S_3),1}=\sum_{a:\,\pi_{v,v'}(a)\in S_3}W_{v,a}$. Hence, by
Lemma~\ref{lem:coarse}, the product for $b=1$ is the sum of the $W_{v,a}$
over $a\in S_1\cap S_2=\emptyset$ (with $\pi_{v,v'}(a)\in S_3$).
For $b=0$, we have 
\begin{align*}
    Q'_{(v,S_1),0}Q'_{(v,S_2),0}Q'_{(v',S_3),0} &= (\id-Q'_{(v,S_1),1})(\id-Q'_{(v,S_2),1})(\id-Q'_{(v',S_3),1})\\
&= \id - Q'_{(v,S_1),1} - Q'_{(v,S_2),1} - Q'_{(v',S_3),1} + Q'_{(v,S_1),1}Q'_{(v',S_3),1} + Q'_{(v,S_2),1}Q'_{(v',S_3),1}\\
    &= \sum_{a\in A_v} W_{v,a} - \sum_{a\in S_1} W_{v,a} - \sum_{a\in S_2} W_{v,a} - \sum_{a\in A_v:\pi_{v,v'}(a)\in S_3}W_{v,a} \\&+ \sum_{a\in S_1:\pi_{v,v'}(a)\in S_3}W_{v,a} + \sum_{a\in S_2:\pi_{v,v'}(a)\in S_3}W_{v,a}\\
    &= \sum_{a\in A_v\setminus(S_1\cup S_2) : \pi_{v,v'}(a)\not\in S_3} W_{v,a}
\end{align*}
and this last sum ranges over the empty set 
by the definition of the
hyperedges.
\end{proof}

\subsection{Soundness}\label{subsec:soundness}

The following lemma is a corollary of a result in~\cite{dinur2005hardness} based on the chromatic number of Schrijver graphs, and it is also used
in~\cite[Appendix~C, Step~4]{fei2026hardness}.

\begin{lemma}\label{lem:kneser-colour}
Let $c\geq2$, let $R$ be a finite set with $m=|R|\geq6c$, let
$r=\lceil(m-c)/2\rceil$, and consider a function $F\colon\binom{R}{r}\to[c]$. Then there is
a colour $c_0\in[c]$ such that
\begin{enumerate}[label=(\arabic*)]
\item $|F^{-1}(c_0)|\geq m^{-c-1}\binom{m}{r}$; and
\item $F^{-1}(c_0)$ contains two disjoint sets.
\end{enumerate}
\end{lemma}

\begin{proof}
Let $\mu=m-2r$.
Since $\mu\in\{c-1,c\}$, $F$ colours the Kneser graph $KG_{m,r}$ with at
most $\mu+1$ colours. By~\cite[Corollary~2.3]{dinur2005hardness}, there is
a colour $c_0$ satisfying~(2) such that $F^{-1}(c_0)$ contains at least a
$2/\bigl((\mu+1)\binom{m}{\mu}\bigr)$ fraction of $\binom{R}{r}$. Since
$\mu\leq c$, we have
$\binom{m}{\mu}\leq m^\mu\leq m^c$. Hence,
\[
\frac{2}{(\mu+1)\binom{m}{\mu}}\;\geq\;\frac{2}{(c+1)\,m^c}
\;=\;\frac{2m}{c+1}\cdot m^{-c-1}\;\geq\;m^{-c-1},
\]
where the last inequality holds because $2m\geq12c\geq c+1$. This
gives~(1).
\end{proof}

We also need the next result.

\begin{lemma}[{\cite[Lemma~C.3]{fei2026hardness}}]\label{lem:fmw-A3}
Fix a positive integer $c$. Let $R$ be a nonempty finite set with
$|R|\geq6c$, and let $r$ be an integer such that
$|R|/3\leq r\leq2|R|/3$. Suppose $B_1,\dots,B_m\subseteq R$ are a sequence
of subsets such that $|B_i|\leq c$ for each $i\in[m]$. If at least a
$|R|^{-c-1}$ fraction of $r$-element subsets $S\in\binom{R}{r}$ contain none
of the sets $B_1,\dots,B_m$, then there exists an element $a\in R$ contained
in at least a $6^{-c}/\bigl(c(c+1)\ln|R|\bigr)$ fraction of the sets
$B_1,\dots,B_m$.
\end{lemma}

The following theorem uses the same idea as the soundness analysis
of~\cite[Appendix~C]{fei2026hardness} (Steps~4,~5,~7, and~8 therein), which is
carried out for the specific label-cover instances of~\cite[Theorem~C.2]{fei2026hardness}
as part of their NP-hardness reduction. Since we apply it to a different instance (a parallel repetition of the stretched magic square),
we state it for an arbitrary biregular projective instance and, for
completeness, we include the argument, which follows~\cite{fei2026hardness} closely.\footnote{\cite[Appendix~C, Step~6]{fei2026hardness}, which handles colourings that are not proper,
is not needed.}

\begin{theorem}\label{thm:source-soundness}
Let $c\geq2$ and $\ell=2c^2$. Let $\Phi=(X_1\sqcup X_2,E,A,\pi,\phi)$ be a
projective bipartite label-cover instance in which $E$ is a set, $\pi$ is
uniform, every variable in $X_1$ lies in exactly $d_1$ constraints, and
every variable in $X_2$ lies in exactly $d_2$ constraints, for some
$d_1,d_2\in\N$. Suppose that $|A_v|\geq6c$ for every
$v\in\bigcup_i\cX_i$, and let $N=\max_v|A_v|$. If $H_{c,\ell}(\Phi)$ has a
proper $c$-colouring, then
\[
\sat(\Phi)\geq\frac{6^{-c}}{2c^4(c+1)\ln N}.
\]
\end{theorem}

\begin{proof}
Fix a proper $c$-colouring $F$ of $H_{c,\ell}(\Phi)$, and let
$\eta=6^{-c}/\bigl(c(c+1)\ln N\bigr)$; note that $\ln N\geq\ln12>0$.

\emph{Stage 1.} 
For each $v\in\bigcup_i\cX_i$, apply
Lemma~\ref{lem:kneser-colour} to $R=A_v$ and the function $S\mapsto F(v,S)$. This gives
a colour $c_v\in[c]$ such that at least an $|A_v|^{-c-1}$ fraction of the
sets $S\in\binom{A_v}{r_v}$ satisfy $F(v,S)=c_v$, together with disjoint sets
$S_1(v),S_2(v)\in\binom{A_v}{r_v}$ such that
$F(v,S_1(v))=F(v,S_2(v))=c_v$. Let $B(v)=A_v\setminus(S_1(v)\cup S_2(v))$,
so that $1\leq c-1\leq|B(v)|\leq c$. For each $i\in[\ell]$, choose
$c(i)\in[c]$ such that $c_v=c(i)$ for at least a $1/c$ fraction of
$v\in\cX_i$. Since $\ell=2c^2$, there is a colour $c^*$ with $c(i)=c^*$ for
at least $2c$ indices $i$; let $I$ be a set of exactly $2c$ such indices.

\emph{Stage 2.} 
For $v\in\cX_i$ with $i<\ell$, the
neighbours of $v$ in $\cX_{i+1}$ (with
respect to the graph $\cL$) are obtained by replacing $v_i\in X_1$ by
some $x_2\in X_2$ with $(v_i,x_2)\in E$; symmetrically for the neighbours in
$\cX_i$ of $v'\in\cX_{i+1}$. Hence, every vertex of $\cX_i$ has $d_1$
neighbours in $\cX_{i+1}$, and every vertex of $\cX_{i+1}$ has $d_2$
neighbours in $\cX_i$. Let $v^{(1)}$ be chosen uniformly at random in $\cX_1$, and, for
$i\in[\ell-1]$, let $v^{(i+1)}$ be a uniformly random neighbour of $v^{(i)}$ in
$\cX_{i+1}$. 
We claim that $(v^{(i)},v^{(j)})$ is uniformly distributed over $\cE_{i,j}$
for all $i<j$. The step from $v^{(s)}$ to $v^{(s+1)}$ changes only the
$s$-th coordinate, from some $u_s\in X_1$ to some $u'_s\in X_2$ with
$(u_s,u'_s)\in E$. Hence,
\begin{align}
\label{eqn_1959_2510}
v^{(i)}=(u'_1,\dots,u'_{i-1},u_i,\dots,u_{\ell-1})\qquad(i\in[\ell]),
\end{align}
and the walk $(v^{(1)},\dots,v^{(\ell)})$ corresponds to the sequence
$\bigl((u_s,u'_s)\bigr)_{s\in[\ell-1]}\in E^{\ell-1}$; clearly, every element of
$E^{\ell-1}$ arises in this way. Since $v^{(1)}$ is uniform in $\cX_1$ and
every vertex of $\cX_s$ has exactly $d_1$ neighbours in $\cX_{s+1}$, each
walk has probability $|X_1|^{-(\ell-1)}d_1^{-(\ell-1)}=|E|^{-(\ell-1)}$.
Now fix $i<j$. By~\eqref{eqn_1959_2510}, $(v^{(i)},v^{(j)})\in\cE_{i,j}$. Conversely,
for $(v,v')\in\cE_{i,j}$, a walk satisfies $v^{(i)}=v$ and $v^{(j)}=v'$ if,
and only if, $u'_s=v_s$ for $s<i$, $(u_s,u'_s)=(v_s,v'_s)$ for
$i\leq s<j$, and $u_s=v_s$ for $s\geq j$. This leaves $d_2$ choices of
$u_s$ for each $s<i$ and $d_1$ choices of $u'_s$ for each $s\geq j$, so
exactly $d_2^{\,i-1}d_1^{\,\ell-j}$ walks pass through $(v,v')$. As this
number does not depend on $(v,v')$, the claim follows.
In particular, $v^{(i)}$ is uniform in $\cX_i$ for every $i\in[\ell]$:
by~\eqref{eqn_1959_2510}, an outcome satisfies $v^{(i)}=v$ if and only if
$u'_s=v_s$ for $s<i$ and $u_s=v_s$ for $s\geq i$, which leaves $d_2$
choices of $u_s$ for each $s<i$ and $d_1$ choices of $u'_s$ for each
$s\geq i$; i.e., $d_2^{\,i-1}d_1^{\,\ell-i}$ outcomes, independently of $v$.

For $i\in I$, let $\Gamma_i$ be the event $c_{v^{(i)}}=c(i)$. Since
$v^{(i)}$ is uniform in $\cX_i$, $\Pr[\Gamma_i]\geq1/c$. By the inclusion-exclusion principle (in particular, by the second Bonferroni inequality),
\[
1\geq\Pr\Bigl[\bigcup_{i\in I}\Gamma_i\Bigr]
\geq\sum_{i\in I}\Pr[\Gamma_i]-\sum_{\substack{i,i'\in I\\ i<i'}}\Pr[\Gamma_i\cap\Gamma_{i'}]
\geq2-\binom{2c}{2}\max_{\substack{i,i'\in I\\ i<i'}}\Pr[\Gamma_i\cap\Gamma_{i'}].
\]
Therefore, there exist $i<j$ in $I$ such that $\Pr[\Gamma_i\cap\Gamma_j]\geq\binom{2c}{2}^{-1}\geq1/(2c^2)$.
From now on, $i$ and $j$ are fixed.
Since $c(i)=c(j)=c^*$ by definition of $I$ and $(v^{(i)},v^{(j)})$ is uniform over $\cE_{i,j}$,
the set $\cE^*=\{(v,v')\in\cE_{i,j}:c_v=c_{v'}=c^*\}$ satisfies
$|\cE^*|\geq|\cE_{i,j}|/(2c^2)$.

\emph{Stage 3.} 
For $v'\in\cX_j$, let
$N^*(v')=\{v:(v,v')\in\cE^*\}$. Fix $v'$ with $N^*(v')\neq\emptyset$, so
that $c_{v'}=c^*$, and let $B(v,v')=\pi_{v,v'}(B(v))$ for
$v\in N^*(v')$; then $|B(v,v')|\leq c$. If $S\in\binom{A_{v'}}{r_{v'}}$
contains $B(v,v')$ for some $v\in N^*(v')$, then
$\{(v,S_1(v)),(v,S_2(v)),(v',S)\}$ is a hyperedge of $H_{c,\ell}(\Phi)$ by Definition~\ref{def:kneser};
as $F$ is a proper colouring and $F(v,S_1(v))=F(v,S_2(v))=c_v=c^*$, we deduce that
$F(v',S)\neq c^*$. Hence every $S$ with $F(v',S)=c^*=c_{v'}$ contains none
of the sets $B(v,v')$ for $v\in N^*(v')$, and by Stage~1 such sets $S$ form at
least an $|A_{v'}|^{-c-1}$ fraction of $\binom{A_{v'}}{r_{v'}}$. Since
$|A_{v'}|\geq6c$, we have $|A_{v'}|/3\leq r_{v'}\leq2|A_{v'}|/3$, so
Lemma~\ref{lem:fmw-A3}, applied to the sequence $(B(v,v'))_{v\in N^*(v')}$
and combined with $\ln|A_{v'}|\leq\ln N$, yields a label
$\tilde a(v')\in A_{v'}$ that lies in $B(v,v')$ for at least an $\eta$
fraction of $v\in N^*(v')$. For every other $v'\in\cX_j$ (i.e., those for which $N^*(v')=\emptyset$), let
$\tilde a(v')\in A_{v'}$ be arbitrary.

Now choose, for each $v\in\cX_i$, a uniformly random label $\tilde a(v)\in B(v)$. Call a pair $(v,v')\in\cE^*$ \emph{good} if $\tilde a(v')\in B(v,v')$. Since $\cE^*$ is the disjoint union of the sets $N^*(v')\times\{v'\}$ over $v'\in\cX_j$ with $N^*(v')\neq\emptyset$, the choice of the labels $\tilde a(v')$ shows that at least $\eta|\cE^*|$ pairs in $\cE^*$ are good. If $(v,v')$ is good, then some $b\in B(v)$ satisfies $\pi_{v,v'}(b)=\tilde a(v')$, so $\Pr[\pi_{v,v'}(\tilde a(v))=\tilde a(v')]\geq1/|B(v)|\geq1/c$. By linearity of expectation, the expected number of pairs $(v,v')\in\cE^*$ with $\pi_{v,v'}(\tilde a(v))=\tilde a(v')$ is at least $\eta|\cE^*|/c$. Using $|\cE^*|\geq|\cE_{i,j}|/(2c^2)$, we deduce that there exist labels $\tilde a(v)\in A_v$  for $v\in\cX_i\cup\cX_j$ such that
\begin{equation}\label{eq:layer-decoding}
\Pr_{(v,v')\in\cE_{i,j}}\bigl[\pi_{v,v'}(\tilde a(v))=\tilde a(v')\bigr]\geq\frac{\eta}{2c^3}.
\end{equation}

\emph{Stage 4.} 
Choose independently: $y_s=y'_s\in X_2$ uniformly for $s<i$;
$(y_s,y'_s)\in E$ uniformly for $i<s<j$; and $y_s=y'_s\in X_1$ uniformly
for $s\geq j$ (the $i$-th coordinate is left free). For $w\in X_1$ and $w'\in X_2$,
let
\[
\iota_1(w)=(y_1,\dots,y_{i-1},w,y_{i+1},\dots,y_{\ell-1})\in\cX_i,\qquad
\iota_2(w')=(y'_1,\dots,y'_{i-1},w',y'_{i+1},\dots,y'_{\ell-1})\in\cX_j.
\]
If $(w,w')\in E$, then $(\iota_1(w),\iota_2(w'))\in\cE_{i,j}$, and
$\pi_{\iota_1(w),\iota_2(w')}$ acts on the $i$-th coordinate as
$\pi_{ww'}$. Now let $(w,w')$ be uniform in $E$ and independent of the random
choices above. By definition of $\cE_{i,j}$, the map
\begin{equation}\label{eq:Eij-bijection}
(v,v')\mapsto\bigl((v_s)_{s<i},\,((v_s,v'_s))_{i\leq s<j},\,(v_s)_{s\geq j}\bigr)
\end{equation}
is a bijection from $\cE_{i,j}$ onto $X_2^{i-1}\times E^{j-i}\times X_1^{\ell-j}$.
Observe that this map sends $(\iota_1(w),\iota_2(w'))$ to
$\bigl((y_s)_{s<i},(w,w'),((y_s,y'_s))_{i<s<j},(y_s)_{s\geq j}\bigr)$,
whose components are independent and uniform. Hence,
$(\iota_1(w),\iota_2(w'))$ is uniform over $\cE_{i,j}$. Let $f(w)\in A_w$ be the $i$-th
coordinate of $\tilde a(\iota_1(w))$ and $f(w')\in A_{w'}$ the $i$-th
coordinate of $\tilde a(\iota_2(w'))$. Whenever
$\pi_{\iota_1(w),\iota_2(w')}(\tilde a(\iota_1(w)))=\tilde a(\iota_2(w'))$,
we have $\pi_{ww'}(f(w))=f(w')$, i.e., $f$ satisfies the constraint
$(w,w')$. As $\pi$ is uniform, the expectation of $\sat_f(\Phi)$ over the random choices of the $y_s$ and $y'_s$ equals the probability, over these choices and $(w,w')$, that $f$ satisfies $(w,w')$; by the above and~\eqref{eq:layer-decoding}, this is at least $\eta/(2c^3)$. Hence,
$\sat(\Phi)\geq\eta/(2c^3)=6^{-c}/\bigl(2c^4(c+1)\ln N\bigr)$, as required.
\end{proof}

\subsection{Proof of Theorem~\ref{thm:source}}\label{subsec:source-proof}

We can now show the existence of a strongly pseudo-telepathic $3$-uniform hypergraph.

\begin{proof}[Proof of Theorem~\ref{thm:source}]
Let $c,k\in\N$; we may assume without loss of generality that $c\geq2$. Let $\Psi=\Psi_k=\Psi(S_k)$ be the label-cover instance described in Section~\ref{subsec:Psi} with stretching length $\lambda=k$.
By Proposition~\ref{prop:Psi}, $\Psi$ is a
projective bipartite label-cover instance in which $E$ is a set, $\pi$ is
uniform, all variables of $X_1$ (respectively, $X_2$) lie in the same
number of constraints, and $\sat(\Psi)<1$. Moreover, by Proposition~\ref{prop:Psi-compatible}, $\Psi$ admits a perfect
$k$-compatible quantum assignment over 
$\C^4$. By
Theorem~\ref{thm:raz}, there is some $\gamma>0$ such that
$\sat(\Psi^{\otimes n})\leq2^{-\gamma n}$ for every $n\in\N$. 
Let $\ell=2c^2$, and choose $n$
in such a way that
\[
2^{-\gamma n}<\frac{6^{-c}}{2c^4(c+1)\,(\ell-1)\,n\ln4}.
\]
By Proposition~\ref{prop:repetition-assignment},
the instance $\Psi^{\otimes n}$ has a perfect $k$-compatible quantum assignment over $(\C^4)^{\otimes n}$. 
Observe also that every equation $u$ of the linear system $S_k$ has $|s(u)|=4$ satisfying assignments, and
$|\F|=2$; hence, by Definition~\ref{def:repetition}, every
variable of $\Psi^{\otimes n}$ in $X_1^n$ has an alphabet of size $4^n$, and every
variable of $\Psi^{\otimes n}$ in $X_2^n$ has an alphabet of size $2^n$. 

Apply now the layered construction of Section~\ref{subsec:layers} to the instance $\Psi^{\otimes n}$.
A vertex
$v=(v_1,\dots,v_{\ell-1})\in\cX_i$ has $i-1$ coordinates in $X_2^n$ and
$\ell-i$ coordinates in $X_1^n$, and $A_v=A_{v_1}\times\dots\times A_{v_{\ell-1}}$;
therefore, $|A_v|=2^{n(i-1)}4^{n(\ell-i)}=2^{n(2\ell-i-1)}$. Since this
quantity decreases in $i$,  we obtain
$6c\leq2^{2c^2-1}\leq2^{n(\ell-1)}\leq|A_v|\leq4^{n(\ell-1)}=:N$.
We deduce that $\Psi^{\otimes n}$ satisfies the hypotheses of
Theorem~\ref{thm:source-soundness} (see also the discussion after
Definition~\ref{def:repetition}).
If $H=H_{c,\ell}(\Psi^{\otimes n})$ had a proper $c$-colouring,
Theorem~\ref{thm:source-soundness} would give
$\sat(\Psi^{\otimes n})\geq6^{-c}/\bigl(2c^4(c+1)(\ell-1)n\ln4\bigr)>2^{-\gamma n}$, a
contradiction. Therefore, $\chi(H)>c$.
Finally, by
Proposition~\ref{prop:kneser-colouring}, $H$ admits a
perfect $k$-compatible quantum $2$-colouring over
$(\C^4)^{\otimes n(\ell-1)}$. This concludes the proof of the theorem.
\end{proof}

\section{Quantum completeness of the Fei--Minzer--Wang game}\label{sec:fmw}
In this section, we show that the reduction in~\cite{fei2026hardness} is quantum complete and preserves compatibility. We only present the components of the game constructed therein that are needed in the quantisation of the completeness side, and we refer to~\cite{fei2026hardness} for the full description of the construction. 
Whenever possible, we try to stick to the notation of~\cite{fei2026hardness}. We only make the following changes to avoid clashes
with the notation of the previous sections: 
\begin{itemize}
    \item we write
$\kappa$ for the parameter denoted $k$ in~\cite{fei2026hardness} (the number of
triples of a tripled set);
\item we write $\varepsilon_H$ for the hypergraph parameter
denoted $\varepsilon$ in~\cite{fei2026hardness};
\item we write $j^*$ for the critical index
denoted $c$ in~\cite{fei2026hardness};
\item and we write $\Adv_j$ for the $j$-th component of an advice
   sequence, denoted $A_j$ in~\cite{fei2026hardness}.
\end{itemize}
The soundness analysis of~\cite{fei2026hardness} is used
as a black box (Theorem~\ref{thm:fmw-soundness}).

\subsection{The game}\label{subsec:fmw-game}
Let $H$ be a $3$-uniform hypergraph and $\kappa\in\N$. For $i\in[\kappa]$,
let $\sE_i(H,\kappa)=\{\{i\}\times \{v_1,v_2,v_3\}: \{v_1,v_2,v_3\}\in E(H)\}$. Following
\cite[Definition~4.2]{fei2026hardness}, $\sE(H,\kappa)$ is the collection of the sets
$E\subseteq[\kappa]\times V(H)$ obtained by picking one edge from each
$\sE_i(H,\kappa)$, i.e., of the sets $E=\bigcup_{i\in[\kappa]}\{i\}\times  \{v_1^{i},v_2^{i},v_3^{i}\}$
with $\{v_1^{i},v_2^{i},v_3^{i}\}\in E(H)$; 
we call such a set $E$ a \textit{tripled set}, and we let
$\Triples(E)$ be the set of the $\kappa$ sets $\{i\}\times\{v^i_1,v^i_2,v^i_3\}$, $i\in[\kappa]$; note that
they partition $E$.
For $t\in\N$, $\sE(H,\kappa,t)$ is the set of sequences
$\cE=(E_1,\dots,E_t)$ with $E_j\in\sE(H,\kappa)$ for all
$j\in[t]$~\cite[Definition~6.28]{fei2026hardness}.

For a finite set $Z$, the space $\F^Z$ has standard basis $(e_z)_{z\in Z}$,
and for $Z'\subseteq Z$ we regard $\F^{Z'}$ as a linear subspace of the ambient space $\F^Z$
\cite[Remark~4.6]{fei2026hardness}. For a sequence $\mathcal{Z}=(Z_1,\dots,Z_t)$ of
finite sets, $\cV(\mathcal Z)=\bigoplus_{j\in[t]}\F^{Z_j}\otimes\F^{Z_j}$
\cite[Definition~6.3]{fei2026hardness}, and for a sequence $\cU=(U_1,\dots,U_t)$ of
subspaces $U_j\subseteq\F^{Z_j}$, we let
$\oplusU\cU=\bigoplus_{j\in[t]}U_j\otimes U_j\subseteq\cV(\mathcal Z)$
\cite[Definition~6.5]{fei2026hardness}. 
Let $U^*$ denote the dual space of $U$; i.e., the space of all linear functionals $U\to\F$.
As in~\cite[\S3.1]{fei2026hardness}, a linear map
$g\colon U\otimes U'\to\F$ is regarded as a bilinear form, its rank is the
rank of the map $U\to(U')^*$, $u\mapsto g(u\otimes\cdot\,)$, and for $U=U'$
it is symmetric if $g(u\otimes w)=g(w\otimes u)$ for all $u,w\in U$. For
$h\in(\F^Z)^*$, we write $h\otimes h$ for the linear map on
$\F^Z\otimes\F^Z$ given by $x\otimes y\mapsto h(x)h(y)$.

Pick a tripled set $E\in\sE(H,\kappa)$.
Following~\cite[Definition~6.1]{fei2026hardness}, let
$N(E)=N_1(E)+N_2(E)\subseteq\F^E\otimes\F^E$, where $N_1(E)$ is the space spanned by
the vectors
\begin{align}
\label{eqn_1558_2110}
n_T=\sum_{a=1}^{3}e_{(i,v_a)}\otimes e_{(i,v_a)}
+e_{(i,v_1)}\otimes e_{(i,v_2)}+e_{(i,v_2)}\otimes e_{(i,v_3)}
+e_{(i,v_3)}\otimes e_{(i,v_1)}
\end{align}
for $T=\{i\}\times\{v_1,v_2,v_3\}\in\Triples(E)$,
and $N_2(E)=\spn\{x\otimes y+y\otimes x:x,y\in\F^E\}$. A linear map
$N(E)\to\F$ is \emph{SymNAE-satisfying} if it maps every vector $n_T$ to $1$ and
every vector $x\otimes y+y\otimes x$ to $0$. By~\cite[Proposition~6.2]{fei2026hardness},
there is exactly one SymNAE-satisfying map on $N(E)$. 
For a sequence $\cE=(E_1,\dots,E_t)$ of tripled sets and a
sequence $\cU=(U_1,\dots,U_t)$ of subspaces $U_j\subseteq\F^{E_j}$, let
$\cN(\cE,\cU)=\bigoplus_{j\in[t]}\bigl(N(E_j)+U_j\otimes U_j\bigr)\subseteq\cV(\cE)$
\cite[Definition~6.6]{fei2026hardness}. A linear map on $\cN(\cE,\cU)$ is
\emph{SNR-satisfying} if, for every $j\in[t]$, its restriction to $N(E_j)$
is SymNAE-satisfying and its restriction to $U_j\otimes U_j$ has rank at
most $1$~\cite[Definition~6.7]{fei2026hardness}.

For the distribution $\Jin$, the advice collections $\sA_{j^*}(\cdot)$, the
sub-advice $\cA_{[q]}$, and the subspaces $\sigma(\cdot)$, we refer
to~\cite[Definitions~6.9, 6.14, 6.18, and Notation~6.17]{fei2026hardness}, respectively; the
properties that we need are collected in Lemma~\ref{lem:fmw-properties}.

\begin{definition}[{\cite[Definition~6.31]{fei2026hardness}}]\label{def:game3}
Let $\kappa,\ell_1,\ell_2,t,s\in\N$ with $t\geq 2$, $\beta\in[0,1)$, and let $H$ be a
$3$-uniform hypergraph. In the two-prover one-round game
$\GameThree[H,\kappa,\ell_1,\ell_2,t,s,\beta]$, the verifier executes the
following:
\begin{enumerate}[nosep]
\item Sample a uniformly random sequence
  $\cE=(E_1,\dots,E_t)\in\sE(H,\kappa,t)$.
\item Sample $(j^*,\cU,\cE',\cU')\sim\Jin(\cE,\beta,\ell_2)$.
\item Sample a uniformly random sequence
  $\cA=(\Adv_1,\dots,\Adv_t)\in\sA_{j^*}(\cE',2\ell_1,s)$. Let
  $L=\sigma(\cA_{[\ell_1]})\subseteq\cV(\cE')$ and
  $L'=\sigma(\cA_{[\ell_1+1]})\subseteq\cV(\cE')$.
\item Send $(\cE,\cU,L+\cN(\cE,\cU))$ to Alice and
  $(j^*,\cE',\cU',L'+\oplusU\cU')$ to Bob.
\item Alice must return a linear map $f\colon L+\cN(\cE,\cU)\to\F$ such
  that $f|_{\cN(\cE,\cU)}$ is SNR-satisfying; Bob must return a linear map
  $g\colon L'+\oplusU\cU'\to\F$ such that, for every $j\in[t]$,
  $g|_{U'_j\otimes U'_j}$ is symmetric and of rank at most $1$. Accept if
  $g|_{L+\oplusU\cU}=f|_{L+\oplusU\cU}$.
\end{enumerate}
\end{definition}

\begin{lemma}\label{lem:fmw-properties}
For every outcome of Steps~1--3 of Definition~\ref{def:game3}, the
following hold.
\begin{enumerate}[label=(\alph*)]
\item $E'_j=E_j$ for $j\neq j^*$, and $E'_{j^*}\subseteq E_{j^*}$ meets
  every triple of $E_{j^*}$ either in the whole triple or in exactly one
  element.
\item $U_j\subseteq\F^{E'_j}$ for every $j$; $U'_j=U_j$ for $j\neq j^*$;
  and $U'_{j^*}=U_{j^*}+\spn(x)$ for some $x\in\F^{E'_{j^*}}$.
\item $L\subseteq L'\subseteq\cV(\cE')\subseteq\cV(\cE)$. 
\item $\oplusU\cU\subseteq\oplusU\cU'\subseteq\cV(\cE')$ and
  $\oplusU\cU\subseteq\cN(\cE,\cU)$. In particular, $L+\oplusU\cU$ is
  contained both in $L+\cN(\cE,\cU)$ and in $L'+\oplusU\cU'$.
\end{enumerate}
\end{lemma}

\begin{proof}
(a) and (b): By~\cite[Definition~6.9]{fei2026hardness}, the index $j^*\in[t]$ is picked
uniformly at random from $[t]$, and
\[
(E'_{j^*},\empt,\empt,U_{j^*})\sim\Jmid(E_{j^*},0,0,\beta,\ell_2),
\qquad U'_{j^*}=U_{j^*}+\spn(x)\ \text{ for a random } x\in\F^{E'_{j^*}},
\]
while, for $j\neq j^*$,
\[
(E_j^{\mathrm{ghost}},\empt,\empt,U_j)\sim\Jmid(E_j,0,0,\beta,\ell_2),
\qquad E'_j=E_j,\qquad U'_j=U_j.
\]
(Here, $\empt$ denotes the empty sequence of vectors, see~\cite[\S3.1]{fei2026hardness}.)
By~\cite[Definitions~4.5 and~5.1]{fei2026hardness}, a sample
$(E',\empt,\empt,U)\sim\Jmid(E,0,0,\beta,\ell_2)$ consists of
$E'\sim\Thin_\beta(E)$ (see~\cite[Definition~4.4]{fei2026hardness}) and the span $U\subseteq\F^{E'}$ of $\ell_2$ random
vectors of $\F^{E'}$ (those in the list denoted by $A'$ in~\cite[Definition~5.1]{fei2026hardness}). Hence, for $j\neq j^*$, we have
$U_j\subseteq\F^{E_j^{\mathrm{ghost}}}\subseteq\F^{E_j}=\F^{E'_j}$
(cf.~\cite[Remark~5.2]{fei2026hardness}), and
$U_{j^*}\subseteq\F^{E'_{j^*}}$. By~\cite[Definition~4.4]{fei2026hardness}, every
triple of $E_{j^*}$ is either included in $E'_{j^*}$ entirely, or reduced
to one of its elements.

(c) By \cite[Notation~6.17 and Definitions~6.12, 6.14, and~6.16]{fei2026hardness},
$L,L'\subseteq\mathcal V(\mathcal E')$, and $\mathcal V(\mathcal E')\subseteq\mathcal V(\mathcal E)$
since $E'_j\subseteq E_j$ for all $j$ by~(a). For $\nu\in\{\ell_1,\ell_1+1\}$,
\cite[Notation~6.17 and Proposition~6.19]{fei2026hardness} give
$\sigma(\mathcal A_{[\nu]})=\mathrm{Mspan}_{j^*}(\mathcal A_{[\nu]},\mathbb F_2^{\nu})
=\mathrm{Mspan}_{j^*}(\mathcal A,\mathbb F_2^{\nu})$, where
$\mathbb F_2^{\ell_1}\subseteq\mathbb F_2^{\ell_1+1}\subseteq\mathbb F_2^{2\ell_1}$ as in that
proposition. Since $\mathrm{Mspan}_{j^*}(\mathcal A,\cdot)$ is monotone with respect to
inclusion, it follows that $L\subseteq L'$.

(d) follows from (b), \cite[Definitions~6.5 and~6.6]{fei2026hardness}, and (c).
\end{proof}

We view the game of Definition~\ref{def:game3} as a CSP instance in the sense of
Section~\ref{subsec:csp}.

\begin{definition}\label{def:PhiF}
Let $\PhiF=\PhiF(H;\kappa,\ell_1,\ell_2,t,s,\beta)$ be the bipartite
label-cover instance defined as follows. The sets $X_1$ and $X_2$ consist
of the Bob questions $p=(j^*,\cE',\cU',D_p)$, where $D_p=L'+\oplusU\cU'$, and of
the Alice questions $q=(\cE,\cU,D_q)$, where $D_q=L+\cN(\cE,\cU)$, respectively, which are sent to the provers with positive probability. The alphabet $A_p$ associated with Bob's question $p$
consists of those $g\in D_p^*$ such that $g|_{U'_j\otimes U'_j}$ is symmetric
of rank at most $1$ for every $j\in[t]$; the alphabet 
$A_q$ associated with Alice's question $q$ is $\{f\in D_q^*:f|_{\cN(\cE,\cU)}\text{ is SNR-satisfying}\}$. For each
outcome $\rho$ of positive probability of the random choices of the
verifier in Steps~1--3 of Definition~\ref{def:game3}, there is a constraint tuple
$\bx_\rho=(p_\rho,q_\rho)$, formed by the two questions sent, of weight
$\pi(\bx_\rho)=\Pr[\rho]$ and with predicate
\[
\phi_{\bx_\rho}=\bigl\{(g,f)\in A_{p_\rho}\times A_{q_\rho}:\ g|_{L_\rho+\oplusU\cU_\rho}=f|_{L_\rho+\oplusU\cU_\rho}\bigr\},
\]
where $L_\rho$ and $\cU_\rho$ are the objects $L$ and $\cU$ determined by the outcome
$\rho$.
\end{definition}

Observe that, by definition, every question is sent for some outcome $\rho$ with positive
probability; in particular, it lies in some constraint $\bx_\rho$.
Clearly, classical assignments of $\PhiF$ are deterministic strategies for the
provers, so
$\sat(\PhiF)=\val\bigl(\GameThree[H,\kappa,\ell_1,\ell_2,t,s,\beta]\bigr)$. Note that the
alphabets $A_p$ contain $0$, and the alphabets $A_q$ are nonempty by
Lemma~\ref{lem:encodings}(i) below.

\subsection{Registers and supports}\label{subsec:registers}
Let $z$ be a variable in the label-cover instance $\PhiF(H;\kappa,\ell_1,\ell_2,t,s,\beta)$, and let $\mathcal Z=(Z_1,\dots,Z_t)$ be the sequence
of sets in $z$, i.e., $\mathcal Z=\cE$ if $z$ is an Alice question and
$\mathcal Z=\cE'$ if $z$ is a Bob question. The set of \emph{registers} is
$\cR=[t]\times[\kappa]$. For $(j,i)\in\cR$, let
\[
\supp_{j,i}(z)=\{v\in V(H):(i,v)\in Z_j\}.
\]
We shall need the following simple observation.
\begin{lemma}\label{lem:supports}
\begin{enumerate}[label=(\roman*)]
\item If $z$ is an Alice question, then every $\supp_{j,i}(z)$ is a
  hyperedge of $H$; if $z$ is a Bob question, then every $\supp_{j,i}(z)$
  is a hyperedge or a singleton.
\item For every constraint $(p,q)$ of $\PhiF$ and every $(j,i)\in\cR$,
  $\supp_{j,i}(p)\subseteq\supp_{j,i}(q)$.
\item $Z_j=\bigcup_{i\in[\kappa]}\{i\}\times\supp_{j,i}(z)$ for every
  $j\in[t]$.
\end{enumerate}
\end{lemma}

\begin{proof}
Since $E_j\in\sE(H,\kappa)$, we have $E_j\cap(\{i\}\times V(H))=\{i\}\times e_{j,i}$
for a hyperedge $e_{j,i}$; for Bob, $(i)$ follows from
Lemma~\ref{lem:fmw-properties}(a). $(ii)$ holds because $E'_j\subseteq E_j$.
 (iii) is clear from the definition of $\supp_{j,i}(z)$.
\end{proof}

\subsection{The quantum assignment}\label{subsec:fmw-assignment}

Fix $k\in\N$ and a perfect $k$-compatible quantum $2$-colouring
$Q=\{Q_{v,b}\}$ of a $3$-uniform hypergraph $H$ over $\cH$.
We shall build a quantum assignment for $\PhiF$
over the Hilbert space $\cH^{\otimes\cR}$, obtained by taking one
copy of $\cH$ for each register.

For a question $z$ with sequence of sets $\mathcal Z$, let $\Omega(z)$ be the
set of families $\omega=(\omega_{j,i})_{(j,i)\in\cR}$ of maps
$\omega_{j,i}\colon\supp_{j,i}(z)\to\F$. Also, for $\omega\in\Omega(z)$, let
\begin{align}
\label{eqn_1306_2110}
\Pi_z(\omega)=\bigotimes_{(j,i)\in\cR}\ \prod_{v\in\supp_{j,i}(z)}Q_{v,\omega_{j,i}(v)}.
\end{align}

We observe the following.

\begin{lemma}
For each $z\in X_1\cup X_2$,
    $\{\Pi_z(\omega)\}_{\omega\in\Omega(z)}\in\PVM_{\cH^{\otimes\cR}}(\Omega(z))$.
\end{lemma}
\begin{proof}
First, note that $v$ is a vertex of $H$ and $\omega_{j,i}(v)$ is an element of $\F$, so $Q_{v,\omega_{j,i}(v)}$ is a well-defined projector on $\cH$.
Since $Q$ is $k$-compatible, it is in particular $1$-compatible (i.e., oracular). Therefore, by Lemma~\ref{lem:supports}(i), the projectors in
each product of~\eqref{eqn_1306_2110} commute, so each factor is a projector on $\cH$, and $\Pi_z(\omega)$ is a projector on $\cH^{\otimes \mathcal R}$.
Using distributivity of matrix multiplication over matrix sum, we find that, for each register $(j,i)$,
\[
\sum_{\omega_{j,i}\colon\supp_{j,i}(z)\to\F}\ \prod_{v\in\supp_{j,i}(z)}Q_{v,\omega_{j,i}(v)}
=\prod_{v\in\supp_{j,i}(z)}\bigl(Q_{v,0}+Q_{v,1}\bigr)=\id_\cH.
\]
Using now distributivity of tensor product over sum, we deduce that
\begin{align*}
    \sum_{\omega\in\Omega(z)}\Pi_z(\omega)
    &=
    \sum_{\omega\in\Omega(z)}\bigotimes_{(j,i)\in\cR}\ \prod_{v\in\supp_{j,i}(z)}Q_{v,\omega_{j,i}(v)}
    =
    \bigotimes_{(j,i)\in\cR}\
    \sum_{\omega_{j,i}\colon\supp_{j,i}(z)\to\F}\ \prod_{v\in\supp_{j,i}(z)}Q_{v,\omega_{j,i}(v)}\\
    &=
    \bigotimes_{(j,i)\in\cR}\id_\cH=\id_{\cH^{\otimes\mathcal R}}.
\end{align*}
Hence,
the family $\{\Pi_z(\omega)\}_{\omega\in\Omega(z)}$ forms a PVM over $\cH^{\otimes\cR}$ with outcome set $\Omega(z)$, as required.
\end{proof}

For $\omega\in\Omega(z)$ and $j\in[t]$, consider the linear functional $h^\omega_j\in(\F^{Z_j})^*$
given by $h^\omega_j(e_{(i,v)})=\omega_{j,i}(v)$ and extended by linearity; this is well defined by
Lemma~\ref{lem:supports}(iii).
Recall that $\cV(\mathcal Z)=\bigoplus_{j\in[t]}\F^{Z_j}\otimes\F^{Z_j}$, and note that the inclusion $D_z\subseteq\cV(\mathcal Z)$ holds regardless of whether $z$ is an Alice or Bob question. 
We can then define
\[
\enc_z(\omega)=\Bigl(\sum_{j\in[t]}h^\omega_j\otimes h^\omega_j\Bigr)\Big|_{D_z},
\]
where $\sum_jh^\omega_j\otimes h^\omega_j$ is the linear form on
$\cV(\mathcal Z)$ acting as $h^\omega_j\otimes h^\omega_j$ on the $j$-th
summand. For a constraint $(p,q)$ of the instance $\PhiF$ and $\omega\in\Omega(q)$, let
$\omega|_p\in\Omega(p)$ be the family of the restrictions of the maps
$\omega_{j,i}$ to $\supp_{j,i}(p)$, which is well defined by Lemma~\ref{lem:supports}(ii).

\begin{lemma}\label{lem:encodings}
Let $\bx_\rho=(p,q)$ be a constraint of $\PhiF$, and let $\omega\in\Omega(q)$ and $\omega'\in\Omega(p)$.
\begin{enumerate}[label=(\roman*)]
\item If every map $\omega_{j,i}$ is non-constant, then
  $\enc_q(\omega)\in A_q$.
\item The maps $\enc_p(\omega|_p)$ and $\enc_q(\omega)$ agree on
  $L_\rho+\oplusU\cU_\rho$.
\item 
$\enc_p(\omega')\in A_p$.
\end{enumerate}
\end{lemma}

\begin{proof}
(i) Write $h_j=h^\omega_j$. Let $T=\{i\}\times\{v_1,v_2,v_3\}$ be a triple
of $E_j$, and let $b_a=\omega_{j,i}(v_a)$ for $a\in[3]$. Since $b_a^2=b_a$ in $\F$, recalling the definition of $n_T$ in~\eqref{eqn_1558_2110}, we have
\[
(h_j\otimes h_j)(n_T)=b_1+b_2+b_3+b_1b_2+b_2b_3+b_3b_1.
\]
If $\omega_{j,i}$ is non-constant, then the numbers $b_1,b_2,b_3$ are not all equal, and thus the expression above equals $1$.
Moreover,
$(h_j\otimes h_j)(x\otimes y+y\otimes x)=2h_j(x)h_j(y)=0$. Hence, the
restriction of $\enc_q(\omega)$ to $N(E_j)\subseteq D_q$ is
SymNAE-satisfying, and its restriction to $U_j\otimes U_j$ is
$h_j|_{U_j}\otimes h_j|_{U_j}$, of rank at most $1$. It follows that the linear map $\enc_q(\omega)|_{\cN(\cE,\cU)}$ is SNR-satisfying and, thus, $\enc_q(\omega)\in A_q$.

(ii) By Lemma~\ref{lem:fmw-properties}(c,d), $L_\rho+\oplusU\cU_\rho$ is contained
in $D_p\cap D_q$ and in $\cV(\cE')$. On $\F^{E'_j}\otimes\F^{E'_j}$, the
maps $h^\omega_j\otimes h^\omega_j$ and
$h^{\omega|_p}_j\otimes h^{\omega|_p}_j$ agree, because
$h^{\omega|_p}_j=h^\omega_j|_{\F^{E'_j}}$.

(iii) 
For each $j\in[t]$, $U'_j\otimes U'_j\subseteq D_p$ lies in the $j$-th
summand of $\cV(\cE')$, as $U'_j\subseteq\F^{E'_j}$ by Lemma~\ref{lem:fmw-properties}(b). The
restriction of $\enc_p(\omega')$ to $U'_j\otimes U'_j$ is $h^{\omega'}_j|_{U'_j}\otimes h^{\omega'}_j|_{U'_j}$, which is a symmetric form of rank at most $1$. Hence, $\enc_p(\omega')$ belongs to $A_p$.
\end{proof}

\begin{lemma}\label{lem:fmw-marginals}
Let $\bx_\rho=(p,q)$ be a constraint of $\PhiF$.
\begin{enumerate}[label=(\roman*)]
\item For every $\omega'\in\Omega(p)$,
  $\Pi_p(\omega')=\sum_{\omega\in\Omega(q):\,\omega|_p=\omega'}\Pi_q(\omega)$.
\item For every $\omega\in\Omega(q)$ such that $\Pi_q(\omega)\neq O$, every map $\omega_{j,i}$ is
  non-constant.
\end{enumerate}
\end{lemma}

\begin{proof}
(i) Recalling that, by Lemma~\ref{lem:supports}(ii),
$\supp_{j,i}(p)\subseteq\supp_{j,i}(q)$ for every $(j,i)\in\cR$, and using distributivity of
the tensor product over sums, we obtain
\begin{align*}
\sum_{\substack{\omega\in\Omega(q)\\ \omega|_p=\omega'}}\Pi_q(\omega)
&=\sum_{\substack{\omega\in\Omega(q)\\ \omega|_p=\omega'}}\
\bigotimes_{(j,i)\in\cR}\ \prod_{v\in\supp_{j,i}(q)}Q_{v,\omega_{j,i}(v)}\\
&=\bigotimes_{(j,i)\in\cR}\
  \sum_{\substack{\omega_{j,i}\colon\supp_{j,i}(q)\to\F\\
  \omega_{j,i}|_{\supp_{j,i}(p)}=\omega'_{j,i}}}\
\prod_{v\in\supp_{j,i}(q)}Q_{v,\omega_{j,i}(v)}\\
&=\bigotimes_{(j,i)\in\cR}
  \Bigl(\prod_{v\in\supp_{j,i}(p)}Q_{v,\omega'_{j,i}(v)}\Bigr)
  \Bigl(\prod_{v\in\supp_{j,i}(q)\setminus\supp_{j,i}(p)}
  \bigl(Q_{v,0}+Q_{v,1}\bigr)\Bigr)\\
&=\bigotimes_{(j,i)\in\cR}\ \prod_{v\in\supp_{j,i}(p)}Q_{v,\omega'_{j,i}(v)}
=\Pi_p(\omega').
\end{align*}
Here, the third equality holds because
the projectors $Q_{v,b}$ with $v\in\supp_{j,i}(q)$ pairwise commute; the fourth holds
because $Q_{v,0}+Q_{v,1}=\id_\cH$.

(ii) Since a tensor
product of operators vanishes as soon as one of its factors does,
\[
\prod_{v\in\supp_{j,i}(q)}Q_{v,\omega_{j,i}(v)}\neq O
\qquad\text{for every }(j,i)\in\cR.
\]
Fix $(j,i)\in\cR$. By Lemma~\ref{lem:supports}(i),
$\supp_{j,i}(q)=\{v_1,v_2,v_3\}$ is a hyperedge of $H$. If $\omega_{j,i}$ were
constant, with value $b\in\F$, then the product above would equal
$Q_{v_1,b}Q_{v_2,b}Q_{v_3,b}$, which is $O$ because the
quantum strategy $\{Q_{v,b}\}$ is perfect and $1$-compatible
(Section~\ref{subsec:hypergraphs}). Hence, $\omega_{j,i}$ is non-constant.
\end{proof}

We are now ready to define the measurements in the quantum strategy for $\PhiF$.
For a question $z$ and an answer $g\in A_z$, we let
\begin{align}
\label{eqn_1010_2910}
\hat Q_{z,g}=\sum_{\omega\in\Omega(z):\,\enc_z(\omega)=g}\Pi_z(\omega).
\end{align}

\subsection{Compatibility and perfection}\label{subsec:fmw-compatibility}

\begin{proposition}\label{prop:fmw-perfect}
The families $\{\hat Q_{z,g}:g\in A_z\}$  defined in~\eqref{eqn_1010_2910} form a quantum assignment $\hat Q$ for
$\PhiF$ over $\cH^{\otimes\cR}$.
\end{proposition}

\begin{proof}
For a Bob question $p$, the family $\{\hat Q_{p,g}\}_{g\in A_p}$ is the minor (recall Section~\ref{subsec:coarse}) of the PVM
$\{\Pi_p(\omega)\}_\omega$ along $\enc_p$, which takes values in $A_p$ by
Lemma~\ref{lem:encodings}(iii); hence it is a PVM with outcome set $A_p$. For an Alice
question $q$, if $\Pi_q(\omega)\neq O$ then $\enc_q(\omega)\in A_q$ by
Lemmas~\ref{lem:fmw-marginals}(ii) and~\ref{lem:encodings}(i); hence
$\sum_{f\in A_q}\hat Q_{q,f}=\sum_\omega\Pi_q(\omega)=\id$, and
$\{\hat Q_{q,f}\}_{f\in A_q}$ is a PVM with outcome set $A_q$.
\end{proof}

\begin{proposition}\label{prop:fmw-compatible}
The quantum assignment $\hat Q$  is
$k$-compatible, for every choice of the parameters
$\kappa,\ell_1,\ell_2,t,s,\beta$.
\end{proposition}

\begin{proof}
Let $z\neq z'$ be questions with $\dist_{\PhiF}(z,z')=m\leq k$, and let
$z=z_0,z_1,\dots,z_m=z'$ be a path in the Gaifman graph of $\PhiF$. Fix a register $(j,i)$.
For any $\ell\in[m]$, the
consecutive questions $z_{\ell-1},z_\ell$ form
a constraint, so one of them must be an Alice question;
without loss of generality, suppose $z_{\ell-1}$ is an Alice question.
By
Lemma~\ref{lem:supports}, $\supp_{j,i}(z_{\ell-1})$ is a hyperedge of $H$, and it contains $\supp_{j,i}(z_{\ell})$.
In particular, any vertex of
$\supp_{j,i}(z_{\ell-1})$ and any vertex of
$\supp_{j,i}(z_\ell)$ are at distance at most $1$ in the Gaifman graph of
$H$. Choosing a vertex in each $\supp_{j,i}(z_\ell)$, we can then build a lazy walk of length $m$ from 
any
$v\in\supp_{j,i}(z)$ to any $v'\in\supp_{j,i}(z')$; hence,
$\dist_H(v,v')\leq m\leq k$. By $k$-compatibility of the quantum assignment $Q$, the
$(j,i)$-th tensor factors of $\Pi_z(\omega)$ and $\Pi_{z'}(\omega')$
commute, for all $\omega\in\Omega(z)$ and $\omega'\in\Omega(z')$. By
Lemma~\ref{lem:tensor-commute}, $\Pi_z(\omega)$ and $\Pi_{z'}(\omega')$
commute and, thus, so do their minors $\hat Q_{z,g}$ and $\hat Q_{z',g'}$ for any $g\in A_z$ and any $g'\in A_{z'}$. Hence, $\hat Q$ is $k$-compatible.
\end{proof}

\begin{proposition}\label{prop:fmw-perfect_new}
The quantum assignment $\hat Q$ is
perfect, for every choice of the parameters
$\kappa,\ell_1,\ell_2,t,s,\beta$.
\end{proposition}

\begin{proof}
    Let $\bx_\rho=(p,q)$ be a constraint. By Lemma~\ref{lem:fmw-marginals}(i),
$\hat Q_{p,g}=\sum_{\omega\in\Omega(q):\,\enc_p(\omega|_p)=g}\Pi_q(\omega)$.
Hence, the two PVMs $\{\hat Q_{p,g}\}_{g\in A_p}$ and $\{\hat Q_{q,f}\}_{f\in A_q}$ are both minors of the same PVM $\{\Pi_q(\omega)\}_{\omega}$. 
Therefore, by Lemma~\ref{lem:coarse}, 
\begin{align*}
\hat Q_{p,g}\hat Q_{q,f}=\sum_{\substack{\omega\in\Omega(q)\\ \enc_p(\omega|_p)=g,\ \enc_q(\omega)=f}}\Pi_q(\omega).
\end{align*}
If the right-hand side is nonzero, then there is $\omega\in\Omega(q)$ with
$\Pi_q(\omega)\neq O$, $\enc_p(\omega|_p)=g$, and $\enc_q(\omega)=f$. By
Lemma~\ref{lem:encodings}(ii), $g$ and $f$ agree on
$L_\rho+\oplusU\cU_\rho$. By Definition~\ref{def:PhiF}, this means that $(g,f)\in\phi_{\bx_\rho}$. 
By Proposition~\ref{prop:fmw-compatible}, the quantum assignment $\hat Q$ is $k$-compatible and, in particular, $1$-compatible.
Applying Lemma~\ref{lem:perfect}, we deduce that it is perfect.
\end{proof}

\subsection{Soundness}\label{subsec:fmw-soundness}

The following result is~\cite[Lemma~6.33]{fei2026hardness}, with the assumption
\cite[Condition~3.12]{fei2026hardness} spelled out as in~\cite[Table~1]{fei2026hardness}.

\begin{theorem}[{\cite[Lemma~6.33]{fei2026hardness}}]\label{thm:fmw-soundness}
There is an absolute constant $C>0$ with the following property. Let
$\delta\in(0,1/2)$, let $q_0,q_1,q_2,\ell_1,\ell_2,s,r\in\N$, and let
$\vec\tau=(\tau_0,\dots,\tau_{2q_0})\in(0,1)^{2q_0+1}$ satisfy
\begin{gather*}
q_0\geq C\log_2(1/\delta),\quad \tau_{2q_0}\leq\exp(-q_0^C),\quad
\tau_{i-1}\leq C^{-1}\tau_i^C\ \ (i\in[2q_0]),\quad
\ell_1\geq C\log_2(1/\tau_0),\\
s=3\ell_1,\quad q_1=2s(2^{2\ell_1}-2^{2\ell_1-q_0}),\quad q_2\geq C\ell_1^2,\quad
\ell_2\geq Cq_2^2,\quad r=2q_2+1.
\end{gather*}
Let $\varepsilon_H\in(0,1)$ with
$\varepsilon_H\leq\exp(-\exp(\exp(Cr)))$, and let $\kappa,t\in\N$ and
$\beta\in(0,1)$ satisfy
\[
\kappa\geq\varepsilon_H^{-C}\exp(\exp(\ell_2)),\qquad\beta=\kappa^{-3/4},\qquad
t\geq\exp(\kappa^C).
\]
Suppose that, in every $4^r$-colouring of the $3$-uniform hypergraph $H$, at
least an $\varepsilon_H$ fraction of the hyperedges are monochromatic. Then
$\val\bigl(\GameThree[H,\kappa,\ell_1,\ell_2,t,s,\beta]\bigr)\leq\delta$.
\end{theorem}

The conditions on $q_0,\vec\tau,\ell_1,s,q_1,q_2,\ell_2,r$ can be met by
choosing these parameters in the order listed, and then those on
$\varepsilon_H,\kappa,\beta,t$ in the order listed.

\begin{corollary}\label{cor:fmw-soundness}
Let $\delta$, $q_0,q_1,q_2,\ell_1,\ell_2,s,r$, and $\vec\tau$ be as in
Theorem~\ref{thm:fmw-soundness}, and let $H$ be a $3$-uniform hypergraph
with $\chi(H)>4^r$. Then there are $\kappa,t\in\N$ and $\beta\in(0,1)$
such that
\[
\sat\bigl(\PhiF(H;\kappa,\ell_1,\ell_2,t,s,\beta)\bigr)\leq\delta.
\]
\end{corollary}

\begin{proof}
Since $\chi(H)>4^r$, every $4^r$-colouring of $H$ has a monochromatic
hyperedge, i.e., at least a $1/|E(H)|$ fraction of monochromatic
hyperedges. Apply Theorem~\ref{thm:fmw-soundness} with
\[
\varepsilon_H=\min\bigl\{1/|E(H)|,\ \exp(-\exp(\exp(Cr)))\bigr\},
\]
and recall that $\sat(\PhiF)=\val(\GameThree)$.
\end{proof}

\section{Exact $d$-to-$1$ constraints}\label{sec:four-to-one}

It is shown in~\cite{fei2026hardness} that
$\GameThree$ is at most $4$-to-$1$.

\begin{lemma}[\protect{\cite[\S6.3]{fei2026hardness}}]\label{lem:at-most-4to1}
The instance $\PhiF=\PhiF(H;\kappa,\ell_1,\ell_2,t,s,\beta)$ is at most
$4$-to-$1$.
\end{lemma}

The reduction to exactly $4$-to-$1$
games is made in~\cite[Appendix~B]{fei2026hardness}. 
We next show that a modification of that construction applies to each $d\geq 4$ at a small cost in soundness and, moreover, it
preserves perfect quantum assignments.

\begin{lemma}\label{lem:completion}
Let $\Phi=(X_1\sqcup X_2,E,A,\pi,\phi)$ be an at most $d$-to-$1$ bipartite
label-cover instance, let $M=\max_{x\in X_1}|A_x|$, and let $N\in\N$ satisfy
$N\geq\max_{x\in X_2}|A_x|$ and $dN>M$. Then there is a $d$-to-$1$ bipartite
label-cover instance $\Phi^N$ with variables $X_1\sqcup X_2$, in which every
variable in $X_1$ has alphabet $A_1$ and every variable in $X_2$ has
alphabet $A_2$, where $|A_1|=dN$ and $|A_2|=N$, such that
\begin{enumerate}[label=(\roman*)]
\item $\Phi^N$ and $\Phi$ have the same Gaifman graph;
\item $\sat(\Phi^N)\leq\sat(\Phi)+d/(dN-M)$;
\item for $k\geq 1$, if $\Phi$ admits a perfect $k$-compatible quantum assignment over $\cH$, then the same is true for $\Phi^N$.
\end{enumerate}
\end{lemma}

\begin{proof}
Let $A_1=[dN]\times\{1\}$ and $A_2=[N]\times\{2\}$, and fix injective maps
$\iota_x\colon A_x\to A_1$ for $x\in X_1$ and $\iota_x\colon A_x\to A_2$
for $x\in X_2$. For $\bx=(x_1,x_2)\in E$, let $\phi_\bx$ be the graph of
$\pi_\bx\colon D_\bx\to A_{x_2}$ with $D_\bx\subseteq A_{x_1}$ as in Section~\ref{subsec:csp}, let
$D'_\bx=\iota_{x_1}(D_\bx)$, and let $\xi_\bx\colon D'_\bx\to A_2$ be
given by $\xi_\bx(\iota_{x_1}(a))=\iota_{x_2}(\pi_\bx(a))$; note that the fibres of $\xi_\bx$
have at most $d$ elements. Let $\Comp(\bx)$ be the set of maps
$\psi\colon A_1\to A_2$ that extend $\xi_\bx$ and all of whose fibres
have exactly $d$ elements. Since
$|A_1\setminus D'_\bx|=dN-|D'_\bx|=\sum_{b\in A_2}\bigl(d-|\xi_\bx^{-1}(b)|\bigr)$,
the set $\Comp(\bx)$ is nonempty. The instance $\Phi^N$ has, for every
constraint $\bx\in E$ (with multiplicity) and every $\psi\in\Comp(\bx)$,
the constraint tuple $\bx$, with weight $\pi(\bx)/|\Comp(\bx)|$ and
predicate the graph of $\psi$. Hence, $\Phi^N$ is $d$-to-$1$, and (i) holds
by construction.

(ii) Let $f^N$ be a classical assignment for $\Phi^N$, and define a classical
assignment $f$ for $\Phi$ by $f(x)=\iota_x^{-1}(f^N(x))$ if
$f^N(x)\in\iota_x(A_x)$, and arbitrarily otherwise. Fix $\bx=(x_1,x_2)\in E$,
let $b_1=f^N(x_1)$ and $b_2=f^N(x_2)$, and let $p_\bx$ be the probability,
for $\psi$ uniform in $\Comp(\bx)$, that $\psi(b_1)=b_2$; thus
$\sat_{f^N}(\Phi^N)=\mathbb{E}_{\bx\sim\pi}\,p_\bx$. If $b_1\in D'_\bx$, say $b_1=\iota_{x_1}(a_1)$ with $a_1\in D_\bx$, then
$p_\bx=1$ if $\xi_\bx(b_1)=b_2$, in which case $f(x_1)=a_1$ and
$f(x_2)=\pi_\bx(a_1)$, so $f$ satisfies $\bx$; and $p_\bx=0$ otherwise. If $b_1\in A_1\setminus D'_\bx$, then, by the symmetry
of $\Comp(\bx)$ under permutations of $A_1\setminus D'_\bx$,
$p_\bx=\bigl(d-|\xi_\bx^{-1}(b_2)|\bigr)/|A_1\setminus D'_\bx|\leq d/(dN-M)$.
Hence, $\sat_{f^N}(\Phi^N)\leq\sat_f(\Phi)+d/(dN-M)$.

(iii) Let $Q$ be a perfect $k$-compatible quantum assignment for $\Phi$ over
$\cH$, and let $Q^N_{x,\iota_x(a)}=Q_{x,a}$ for $x\in X_1\sqcup X_2$ and
$a\in A_x$, and $Q^N_{x,b}=O$ for all other labels $b$. For $x\in X_1$
(respectively, $x\in X_2$), $Q^N(x)$ is a PVM with outcome set $A_1$ (respectively,
$A_2$). By (i), commutation relations transfer from $Q$ to $Q^N$, so $Q^N$ is $k$-compatible (and, in particular, $1$-compatible). By
Lemma~\ref{lem:perfect}, it is then enough to show that
$Q^N_{x_1,b_1}Q^N_{x_2,b_2}=O$ for every constraint of $\Phi^N$ coming from
$\bx=(x_1,x_2)\in E$ and $\psi\in\Comp(\bx)$, and every pair
$(b_1,b_2)\in A_1\times A_2$ outside the graph of $\psi$. This is clear
unless $b_1=\iota_{x_1}(a_1)$ and $b_2=\iota_{x_2}(a_2)$ for some
$a_1\in A_{x_1}$ and $a_2\in A_{x_2}$. In this case,
$(a_1,a_2)\notin\phi_\bx$, as otherwise
$\psi(b_1)=\xi_\bx(b_1)=\iota_{x_2}(\pi_\bx(a_1))=b_2$. Since $Q$ is perfect,
Lemma~\ref{lem:perfect} gives
$O=Q_{x_1,a_1}Q_{x_2,a_2}=Q^N_{x_1,b_1}Q^N_{x_2,b_2}$.
\end{proof}

We point out that an alternative approach for extending the pseudo-telepathy result from the $4$-to-$1$ case 
to the $d$-to-$1$ case for $d>4$ would have been to establish quantum completeness of the reduction in~\cite[\S4]{guruswami_et_al:LIPIcs.ICALP.2020.62} from exact $d$-to-$1$ to exact $(d+1)$-to-$1$ label-cover games.

\section*{Acknowledgements}
The mathematical ideas and the overall approach presented in this paper are due to the authors. We used Claude Opus~5.5 (Anthropic) to assist our navigation of the technical machinery of~\cite{fei2026hardness}, to
generate lower-level details of the proofs, and to proofread the
manuscript. The authors take full responsibility for
the correctness of the paper.

The first author acknowledges funding from the Steiermärkischen Landesregierung, Referat Wissenschaft und Forschung [grant ref. no. ABT12-77337/2026-6].
The second author acknowledges funding from the project “Hamburg
Quantum Computing”, co-financed by ERDF of the European Union and by the Fonds of the Hamburg
Ministry of Science, Research, Equalities and Districts (BWFGB).

\newpage

\addcontentsline{toc}{section}{References}
\printbibliography

\end{document}